\documentclass[
 reprint,
 amsmath,amssymb,
 aps,
]{revtex4-1}

\usepackage{graphicx}
\usepackage{dcolumn}
\usepackage{bm}
\usepackage{hyperref}
\usepackage{float}
\usepackage{color}
\usepackage{amsmath}
\usepackage{amssymb}
\usepackage{wasysym}
\usepackage{multirow}
\usepackage[normalem]{ulem}
\usepackage{upgreek}
\usepackage{braket}
\usepackage{siunitx}

\newcommand{\opd}[2]{\mbox{$\hat{#1}_{#2}^{\dagger}$}} 
\newcommand{\op}[2]{\mbox{$\hat{#1}_{#2}$}}

\newcommand{\cout}{\op c{\text{out}}}
\newcommand{\nth}{n_\text{th}}
\newcommand{\ncoh}{n_\text{coh}}

\newcommand{\omegao}{\omega_\text{o}}
\newcommand{\omegam}{\omega_\text{m}}
\newcommand{\Zeff}{Z_\text{eff}}
\newcommand{\gammamu}{\gamma_{\upmu}}
\newcommand{\gammai}{\gamma_\text{i}}
\newcommand{\gammaom}{\gamma_\text{om}}
\newcommand{\Gm}{G_\text{m}}
\newcommand{\Go}{G_\text{o}}
\newcommand{\go}{g_\text{o}}
\newcommand{\gmu}{g_\upmu}
\newcommand{\nc}{n_\text{a}}
\newcommand{\etamu}{\eta_{\upmu\text{m}}}
\newcommand{\etasys}{\eta_\text{sys}}
\newcommand{\etao}{\eta_\text{o}}
\newcommand{\etademod}{\eta_\text{d}}
\newcommand{\etaherald}{\eta_\text{herald}}

\newcommand{\kappao}{\kappa_\text{o}}
\newcommand{\kappamu}{\kappa_\upmu}
\newcommand{\kappaoe}{\kappa_\text{o,e}}
\newcommand{\kappamue}{\kappa_{\upmu,\text{e}}}
\newcommand{\nex}{n_\text{ex}}
\newcommand{\Hom}{\op H{\text{om}}}
\newcommand{\Hpe}{\op H{\text{pe}}}
\newcommand{\Hbs}{\op H{\text{bs}}}
\newcommand{\Htms}{\op H{\text{tms}}}
\newcommand{\mathcalD}{\mathcal{D}}

\begin{document}

\preprint{APS/123-QED}

\title{Optically heralded microwave photons}

\author{Wentao Jiang}
\thanks{These authors contributed equally}
\author{Felix M. Mayor}
\thanks{These authors contributed equally}
\author{Sultan Malik}
\author{Rapha\"el Van Laer}
\thanks{Present address: Department of Microtechnology and Nanoscience, Chalmers University of Technology, Sweden}
\author{Timothy P. McKenna}
\author{Rishi N. Patel}
\author{Jeremy D. Witmer}
\author{Amir H. Safavi-Naeini}
\email{safavi@stanford.edu}

\affiliation{Department of Applied Physics and Ginzton Laboratory, Stanford University, 348 Via Pueblo Mall, Stanford, California 94305, USA}

\date{\today}

\begin{abstract}
A quantum network that distributes and processes entanglement would enable powerful new computers and sensors~\cite{kimble2008quantum, altman2021quantum, alexeev2021quantum, awschalom2021development}. Optical photons with a frequency of a few hundred terahertz are perhaps the only way to distribute quantum information over long distances~\cite{chen2021twin}. Superconducting qubits on the other hand, which are one of the most promising approaches for realizing large-scale quantum machines~\cite{devoret2013superconducting, arute2019quantum, wu2021strong, jurcevic2021demonstration}, operate naturally on microwave photons that have roughly $40,000$ times less energy. To network these quantum machines across appreciable distances, we must bridge this frequency gap and learn how to generate entanglement across widely disparate parts of the electromagnetic spectrum. Here we implement and demonstrate a transducer device that can generate entanglement between optical and microwave photons, and use it to show that by detecting an optical photon we add a single photon to the microwave field. We achieve this by using a gigahertz nanomechanical resonance as an intermediary, and efficiently coupling it to optical and microwave channels through strong optomechanical and piezoelectric interactions~\cite{mirhosseini2020superconducting}. We show continuous operation of the transducer with $5\%$ frequency conversion efficiency, and pulsed microwave photon generation at a heralding rate of $15$ hertz.
Optical absorption in the device generates thermal noise of less than two microwave photons.
Joint measurements on optical photons from a pair of transducers would realize  entanglement generation between distant microwave-frequency quantum nodes~\cite{krastanov2021optically}. Improvements of the system efficiency and device performance, necessary to realize a high rate of entanglement generation in such networks are within reach.
\end{abstract}

\maketitle

Manipulating and transmitting quantum states with higher fidelity and at larger scales will enable technologies that promise breakthroughs in sensing, communication, and computation~\cite{altman2021quantum, alexeev2021quantum, awschalom2021development}. Over the last two decades, our ability to manipulate the states of photons in superconducting circuits has advanced rapidly and led to demonstrations of quantum advantage for certain computational tasks~\cite{arute2019quantum,wu2021strong, jurcevic2021demonstration}. Separately, the first quantum networks have been realized based on the transmission of quantum states and distribution of entanglement over a small number of nodes using optically coupled qubits~\cite{pompili2021realization, daiss2021quantum}. Optical interconnects between superconducting quantum machines with advanced computational and error correction capabilities~\cite{ofek2016extending,krinner2022realizing} would significantly accelerate the development and deployment of quantum networks. Moreover, microwave quantum processors  would be beneficiaries of quantum networks that support distributed quantum computing~\cite{serafini2006distributed}, sensing~\cite{gottesman2012longer, brady2022entangled}, and secured communications~\cite{shor2000simple}. Unfortunately, in contrast to ions, atoms, and semiconductor defect centers, superconducting circuits lack a natural optical transition that would generate entanglement between propagating optical photons and their internal states. To compensate for this deficiency, a highly efficient and low-noise quantum transducer needs to be developed.

Quantum transducers that connect microwave and optical photons have used a variety of physical processes including optomechanical, electro-optical, magneto-optical interactions, and atomic degrees of freedom~\cite{han2021microwave}. In optomechanical transducers, the interaction between light and motion via radiation pressure and electrostriction offers the nonlinearity necessary for frequency conversion~\cite{aspelmeyer2014cavity}. The mechanical vibration is then  converted to a microwave signal with either a parametric electromechanical coupling~\cite{teufel2011circuit, andrews2014bidirectional} or the piezoelectric effect~\cite{o2010quantum, bochmann2013nanomechanical}. Steady progress in the field has led to constant improvement of transduction efficiencies and added noise~\cite{zeuthen2020figures, han2020cavity, brubaker2022optomechanical}, culminating in remarkable recent demonstrations of superconducting qubit state readout~\cite{mirhosseini2020superconducting, delaney2022superconducting}. 

Here we split optical photons into correlated pairs of optical and microwave photons in a quantum transducer and herald microwave photons by detecting optical photons. Our demonstration is an important step towards realizing heralded entanglement generation between distant superconducting quantum machines that may form nodes in a quantum network~\cite{kimble2008quantum}. We send a laser pulse with the sum frequency of the optical and microwave frequencies to the transducer. Through spontaneous down-conversion, the input optical photon is converted to an entangled pair consisting of an optical photon and a microwave phonon. The phonon half of this pair efficiently radiates from the device into an output line as a microwave-frequency electromagnetic signal, generated by a strong engineered piezoelectric coupling. The light is sent onto a single photon detector and upon detection, we herald the microwave photon. Finally, we characterize the state of the microwave field using linear detection tomography and verify the presence of the additional photon.

\begin{figure*}[!htbp]
\centering
\includegraphics[scale=1]{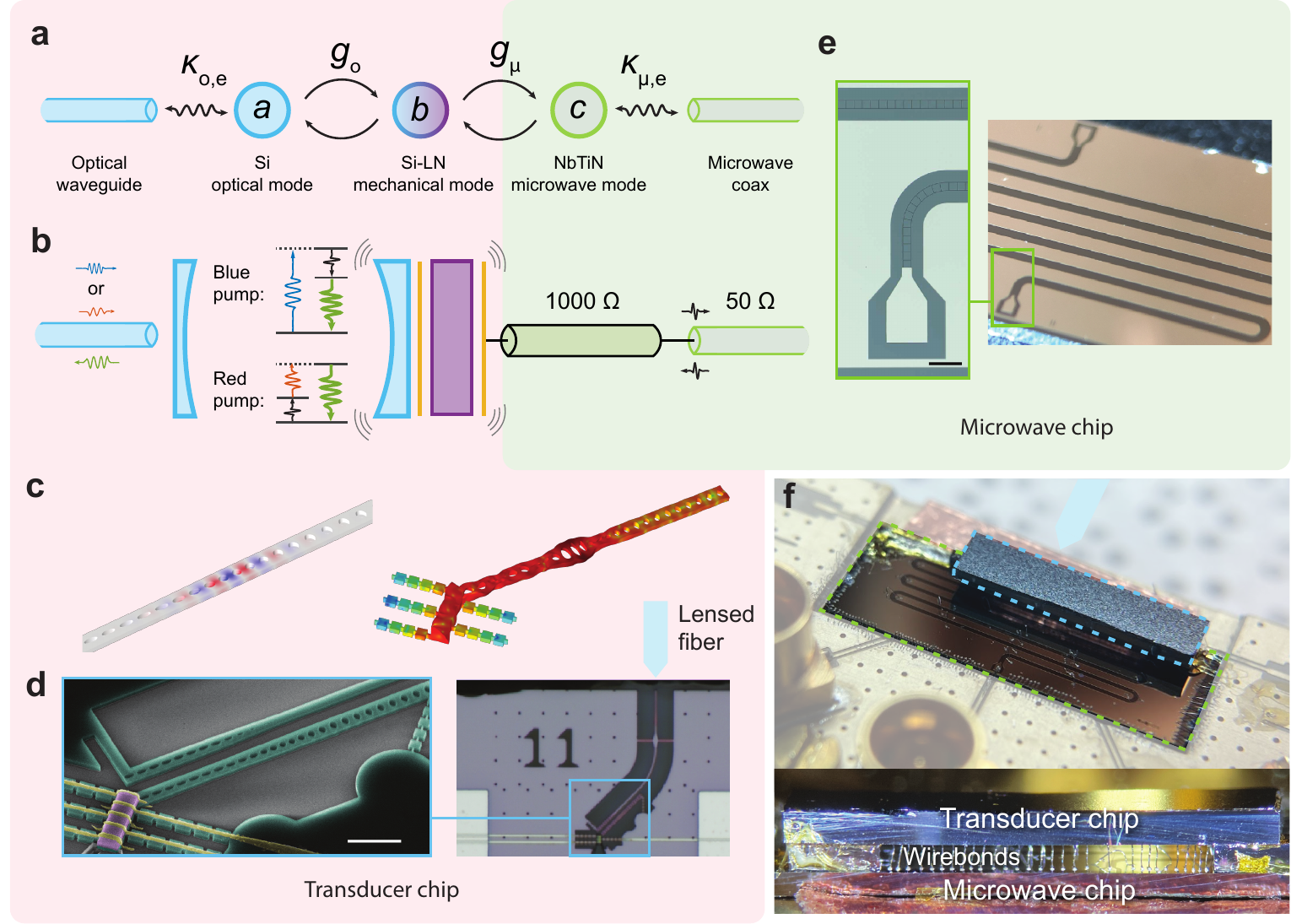}
\caption{\label{Fig1:design} \textbf{Transducer schematics and implementation.} \textbf{a}, Definition of modes in the transducer and the coupling rates between the modes and channels. \textbf{b}, Physical implementation of the coupling mechanism and the possible conversion processes. The silicon optical mode is represented by the blue Fabry-Pérot cavity, and the optomechanical coupling offers two types of three-wave mixing processes between photons in the cavity and phonons in the vibrating mirror. A blue (red) pump laser-cavity detuning allows entangled pair generation (quantum frequency conversion). The hybrid silicon-lithium niobate (Si-LN) mechanical mode utilizes the piezoelectric effect to convert the mechanical vibration to a microwave field. A niobium-titanium-nitride (NbTiN) microwave resonance formed by the standing wave in a high-impedance waveguide enhances the external coupling rate of the mechanical mode to the microwave coaxial cable. \textbf{c}, Simulated optical mode and mechanical mode of the transducer. \textbf{d},\textbf{e}, Optical and scanning electron microscope (SEM) images of the fabricated transducer device and microwave resonator. SEM image of the transducer is taken before etching the silica substrate and false-colored to highlight the silicon (cyan), LN (purple), and aluminum electrodes (yellow). Scale bars: \SI{2}{\micro\meter} (white), \SI{200}{\micro\meter} (black). \textbf{f}, Top: photo of the transducer assembly wirebonded to a printed circuit board. The transducer (microwave) chip is highlighted with dashed blue (green) lines. The lensed fiber is also shown as the light blue region (not to scale) Bottom: side view of the transducer assembly as seen by the lensed fiber, showing the two chips and the wirebonds connecting them.}
\end{figure*}

The two physical processes in the transducer are the nonlinear optomechanical interaction between the optical mode and the mechanical mode, and the linear piezoelectric interaction between the mechanical mode and the microwave mode (Fig.~\ref{Fig1:design}a). The nonlinearity that facilitates the frequency conversion process is governed by the optomechanical interaction Hamiltonian $ \Hom = \hbar \go \opd a{} \op a{} (\op b{} + \opd b{}) $, where $\op a{}$ ($ \op b{}$) is the annihilation operator for the optical (mechanical) resonance. The optomechanical coupling rate $\go$ represents the cavity frequency uncertainty from the zero-point motion in the mechanical mode~\cite{aspelmeyer2014cavity}. As shown in Fig.~\ref{Fig1:design}b, when the system is pumped with a red-detuned laser, the interaction can be described by the beam-splitter Hamiltonian $ \Hbs = \hbar \Go (\op a{} \opd b {} + \opd a {} \op b {})$, while a blue-detuned laser implements a two-mode squeezing Hamiltonian $ \Htms = \hbar \Go (\opd a{} \opd b {} + \op a {} \op b {})$. We define the linearized coupling rate $ \Go = \sqrt{\nc} \go $ where $\nc$ is the intracavity pump photon number. The operator $\op a {}$ now represents the sideband component of the optical mode that is resonant with the cavity. The piezoelectric interaction between the mechanical and microwave resonance is described by $\Hpe = \hbar \gmu (\op b {} \opd c{} + \opd b {} \op c {}) $, where $\gmu $ is the coupling rate and $\op c{}$ is the lowering operator of the microwave mode.

We operate our transducer in a fast-cavity limit, where the optical (microwave) linewidth $\kappao $ ($\kappamu$) is much larger than the corresponding coupling rate $ \kappao \gg \Go$ ($ \kappamu \gg \gmu $). In this limit, both the optical and microwave subsystems act approximately as a broad Markovian bath for the mechanical mode. This means that the resonant phonons decay at a rate $\gammamu = 4\gmu^2/\kappamu$ into the microwave transmission line. Similarly, for a red-side pump, the phonons decay directly into the photon loss channels at a rate of $\gammaom = 4\Go^2/\kappao$.  The peak conversion efficiency is given by the ratio between external coupling rates and the total loss rates,
\begin{equation}
    \eta = \etao \eta_\upmu \frac{4 \gammaom \gammamu}{(\gammai + \gammaom + \gammamu)^2},
\end{equation}
where $\gammai$ is the intrinsic loss rate of the mechanical mode, and $\etao \equiv \kappaoe /\kappao$ ($\eta_\upmu \equiv \kappamue /\kappamu$) is the external coupling efficiency of the optical (microwave) mode (Fig.~\ref{Fig1:design}a). For a blue-side pump, the interaction induces a non-degenerate parametric amplification rate of $\gammaom$ for the mechanical resonator, with correlated photons emitted at optical frequency.

To achieve higher conversion efficiency and bandwidth, larger $\gammaom$ and $\gammamu$ are required. This translates to higher optomechanical and piezoelectric coupling coefficients. We found that no single material system is optimal with respect to all the needs of the converter. As such, we pursued a heterogeneous integration approach that combines materials with good optomechanical and piezoelectric properties. In addition to the materials integration challenge, this raises a challenge in design, as we must co-design the optomechanical and piezoelectric constituents of the transducer to maximize the modal overlaps while maintaining small mode volumes for high interaction rates.  
We implement the transducer by combining highly piezoelectric thin-film lithium niobate (LN)~\cite{pop2017laterally}, with thin-film silicon (Si), which has been shown to have strong optomechanical coupling and low loss~\cite{safavi2019controlling}. We use a silicon optomechanical crystal (OMC) to co-localize the optical and mechanical resonances in a wavelength-scale volume~\cite{chan2012optimized}, and engineer the mechanical mode to be partially extended and strongly hybridized with a Si-LN hybrid piezoelectric mode~\cite{mirhosseini2020superconducting}. The orientation between the piezoelectric resonator and the silicon OMC is chosen to maximize the mechanical hybridization. The full transducer structure is released and supported by one-dimensional silicon phononic shields to minimize unwanted mechanical loss (Fig.~\ref{Fig1:design}c,d). 

To couple the phonons to microwaves, we pattern aluminum electrodes on the LN and run these over the phononic shields that suspend the transducer. Because of the vastly different dimensions of the piezo-optomechanical element ($\sim\SI{10}{\micro\meter}$) and the microwave circuit ($\sim\SI{10}{\milli\meter}$), we fabricate them on separate chips that we combine by wirebonding (Fig.~\ref{Fig1:design}f). The microwave resonance is formed by a standing wave in a high-impedance (high-Z) microwave coplanar waveguide (Fig.~\ref{Fig1:design}b,e) arising due to impedance mismatch with the output line. A niobium-titanium-nitride (NbTiN) thin film on high resistivity silicon substrate is patterned into nanowires to support traveling waves with characteristic impedance $Z = \SI{1000}{\ohm}$.  The large kinetic inductance from the nanowires enables high impedance and magnetic frequency tunability~\cite{xu2019frequency}. Finally, by using NbTiN, with its short quasi-particle lifetime, for the microwave resonator, and fabricating the microwave subsystem on a separate chip, we mitigate some of the effects associated with absorption of stray optical radiation~\cite{mirhosseini2020superconducting}.

\begin{figure*}[!htbp]
\centering
\includegraphics[scale=1]{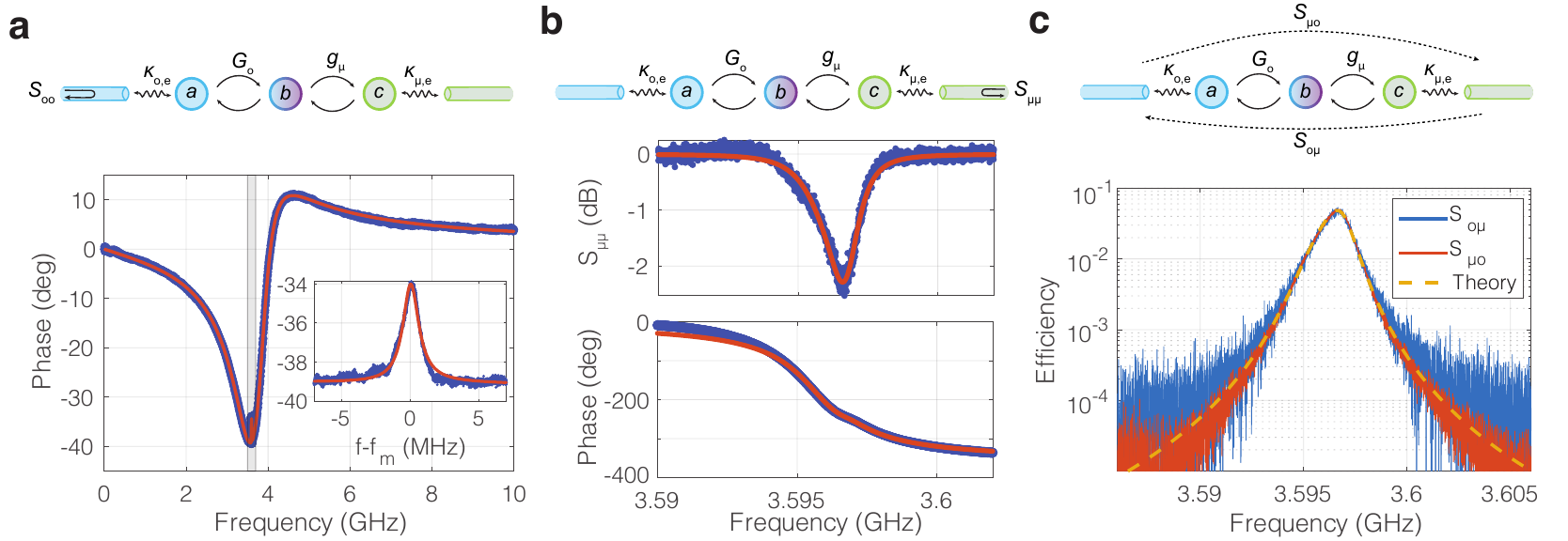}
\caption{\label{Fig2:S-param} \textbf{Transducer characterization.} \textbf{a}, Electromagnetically induced transparency (EIT). Top: schematic of the measurement allowing us to extract the optical reflection $S_\text{oo}$ with a red detuned pump. Bottom: measurement of the phase of the reflected optical sideband (blue) with the EIT visible at the mechanical frequency $f_\text{m} =  \SI{3.596}{\giga\hertz}$. A close-up of the transparency region is shown in the inset. The red line is a fit assuming the model shown in the schematic. \textbf{b}, Microwave reflection. Top: schematic of the measurement of $S_{\upmu\upmu}$. Bottom: Magnitude and phase of the measured $S_{\upmu\upmu}$. The fit (red) is using coupled mode theory. \textbf{c}, Frequency conversion. Schematic and measurement of the microwave-to-optical ($S_{\text{o}\upmu}$) and optical-to-microwave ($S_{\upmu\text{o}}$) conversion efficiencies using a red-detuned pump. The theoretical calculation is shown as yellow dashed line.}
\end{figure*}

We characterize the transducer at the mixing chamber plate in a dilution refrigerator where the temperature $T\lesssim \SI{10}{\milli\kelvin}$ --  the same environment in which superconducting circuits and qubits are operated. We use a lensed fiber to focus light into an on-chip photonic waveguide, to which the optical cavity is evanescently coupled. We first measure the linear scattering parameters of the transducer with a red-detuned continuous-wave pump. We use \SI{4.4}{\micro\watt} on-chip pump power, corresponding to an intracavity photon number $\nc = 230$. We sweep a weak probe tone across the optical resonance, generated by electro-optic modulation of the pump light by the microwave signal from a vector network analyzer (VNA). The reflected light is amplified and detected by a high speed photodetector and subsequently the VNA. We obtain the scattering parameter $S_\text{oo}$ and extract the phase response (Fig.~\ref{Fig2:S-param}a). We observe electromagnetically induced transparency (EIT)~\cite{aspelmeyer2014cavity} and fit the response using input-output theory which gives a single-photon optomechanical coupling rate of $\go/2\pi = \SI{410}{\kilo\hertz}$. The mechanical mode has a frequency $\omegam/2\pi = \SI{3.596}{\giga\hertz}$. We find that the optical cavity is nearly critically coupled with a total linewidth $\kappa_\text{o}/2\pi = \SI{1.12}{\giga\hertz}$. Subsequently, we characterize the piezoelectric interaction in the transducer by a microwave reflection measurement. We obtain the scattering parameter $S_{\upmu\upmu}$ and plot its magnitude and phase (Fig. \ref{Fig2:S-param}b). We find a piezoelectric coupling rate $g_\upmu/2\pi = \SI{420}{\kilo\hertz}$ and extract an intrinsic mechanical linewidth of $\gammai/2\pi= \SI{1.1}{\mega\hertz}$. During this measurement, the pump is sent to the device with \SI{10.1}{\micro\watt} on-chip power ($\gammaom/2\pi = \SI{324}{\kilo\hertz})$.
The device parameters are summarized in Extended Data Table \ref{tab:parameters}. Next, we characterize the microwave-to-optical ($S_{\text{o}\upmu}$) and optical-to-microwave ($S_{\upmu\text{o}}$) frequency conversion using the VNA. The setup is the same as in the reflection measurements with an on-chip pump power of \SI{10.2}{\micro\watt}, corresponding to $\nc = 540$ and $\gammaom/2\pi = \SI{327}{\kilo\hertz}$. The measured scattering parameters are shown in Fig. \ref{Fig2:S-param}c. We determine the peak conversion efficiency to be $\eta = (4.9 \pm 0.5)\%$ (see Methods) with a $\SI{3}{\decibel}$ bandwidth of $\SI{1.5}{\mega\hertz}$. The added noise referred to the input for optical-to-microwave conversion is measured to be $n_{\text{added}, \upmu\text{o}} = 99 \pm 10$. Like state-of-the-art demonstrations, our achieved efficiency and added noise are currently insufficient for direct conversion of quantum states. However, the ability to post-select quantum optical states by single photon detection obviates the need for high efficiency by leveraging the nonlinearity afforded by strong measurement. A more important goal for our device is then to achieve a high rate of entanglement generation so that heralded protocols become practicable.


\begin{figure*}[!htbp]
\centering
\includegraphics[scale=1]{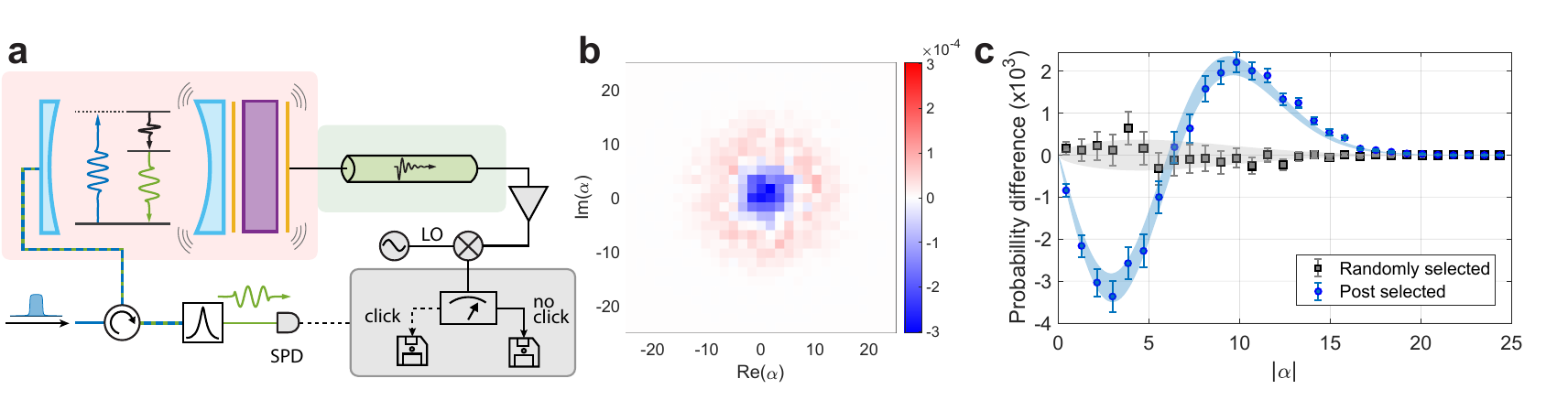}
\caption{\label{Fig3:herald} \textbf{Optically heralded microwave photon.} \textbf{a}, Schematics of the heralding experiment. A blue detuned pump pulse incident on the transducer is split into a sideband photon on-resonance with the optical cavity, and a phonon resonant with the mechanical mode. The reflected light contains the pump residual and the sideband photon, which is filtered and detected with a single photon detector (SPD). The phonon is converted to a propagating microwave photon, and is amplified, downconverted by mixing with a local oscillator (LO), demodulated, and accumulated into a phase space probabilistic distribution $\mathcalD (\alpha)$. When a click is registered on the SPD, the demodulation result is collected into a post-selected dataset $\mathcalD_\text{s}(\alpha)$. \textbf{b}, Probability distribution difference between the post-selected dataset and the full dataset $\mathcalD_\text{s}(\alpha) - \mathcalD(\alpha)$. \textbf{c}, Probability distribution difference of the radially binned distribution. Blue points represent the post-selected samples, and black points are calculated from randomly-selected samples instead of post-selected, with the same number of samples. Errorbars represent one standard deviation according to number of samples in each bin. Blue and grey regions are theoretical calculations with the same uncertainty.}
\end{figure*}

For heralded microwave photon generation, a blue-detuned pump laser pulse with duration $\tau \approx \SI{20}{\nano\second}$ is sent to the transducer. The pulse duration is chosen so that $\tau^{-1}$ is much smaller than the optical decay rate and laser detuning, and so the intracavity photon number follows the pulse amplitude closely. The pulse is however much faster than the response time of the mechanical and microwave system. As shown in Fig.~\ref{Fig3:herald}a, upon heralding, a phonon is effectively added to the mechanical resonator, which then leaks out as microwave radiation at a rate on the order of $\gamma_\upmu$. We integrate the optomechanical scattering rate over the optical pulse duration to find the entangled photon-phonon-pair generation probability $p \approx  \gammaom \tau$~\cite{krastanov2021optically}. The sideband photon leaks out of the optical cavity together with reflected pump photons, most of which are filtered out by two Fabry-Pérot cavities, enabling single photon detection (SPD) of the sideband photon. The generated microwave photon is amplified by a near quantum-limited traveling wave parametric amplifier (TWPA)~\cite{macklin2015near} followed by a cryogenic low-noise amplifier. At the end of a room temperature amplification chain, the microwave signal is demodulated to obtain a quadrature sample. The samples are labeled according to whether the microwave detection was accompanied by a photon click.

We use an on-chip peak pump power of $\sim \SI{5}{\micro\watt}$ and a pulse duration of $\tau \approx \SI{20}{\nano\second}$, giving a scattering probability of $ p \approx 3.6 \%$. We choose a repetition rate of \SI{170}{\kilo\hertz}, which leads to a thermal occupation of the mechanical mode prior to the arrival of the pump pulse of $\nth = 0.68\pm 0.08$ (see Methods). A lower repetition rate reduces this source of noise, but makes the experiment slower. We use a pair of filter cavities to realize  $> \SI{90}{\decibel}$ suppression of the pump photons with respect to the optomechanically generated sideband photons. The probability of detecting a photon that starts inside the optical mode is $\etasys = 1\%$. This low efficiency is caused by the optical mode external coupling efficiency $\etao = 50\%$, an insertion loss of $ 25.4\%$ from the on-chip photonic waveguide to the fridge optical port, transmission through the filter cavities ($15\%$), and quantum efficiency of the SPD ($65\%$).

Finally, we use linear detectors to characterize the microwave field emitted from the device. Linear phase-insensitive amplification of microwave fields effectively measures both quadratures, necessarily adding half a quantum of noise. The resulting amplifier output corresponds to the Husimi $Q$ function $ Q(\alpha) $ of the microwave state up to a scaling~\cite{eichler2011experimental,patel2021room}, where $\alpha$ is the quadrature-phase amplitude of the microwave mode. In practice, noise in excess of the quantum limit is added due to the device inefficiency, amplifier noise, and demodulation inefficiency. We denote the measured probability distribution by $\mathcalD(\alpha) $ for the thermal state without the single-photon detection event, and $\mathcalD_\text{s}(\alpha) $ for the photon added state after post-selection with the SPD event. As a result of the excess noise, the measured $\mathcalD(\alpha) $ and $\mathcalD_\text{s}(\alpha) $ look virtually identical to the eyes but their numerical difference is clearly resolvable~\cite{eichler2011experimental}. We accumulated $\sim 1.4\times 10^6$ post-selected samples over the course of $\sim \SI{35}{\hour} $, and binned them into a two-dimensional histogram with $41\times 41$ entries to obtain the probabilistic distribution $ \mathcalD_\text{s}(\alpha) $. During the same experiment, $ 4.3\times 10^7 $ samples of the thermal state are collected and binned in the same way for $  \mathcalD(\alpha)  $. The difference in the probability distributions $ \mathcalD_\text{s}(\alpha) - \mathcalD(\alpha) $ is shown in Fig.~\ref{Fig3:herald}b. Since the states have no well-defined phase, we bin the samples radially and plot the results as blue points in Fig.~\ref{Fig3:herald}c. The shaded blue region is the resulting theoretical distribution of phonon added state with excess noise. As a control, we randomly sampled a subset from the thermal dataset, and obtained the  probability difference shown in black in Fig.~\ref{Fig3:herald}c. In both the two-dimensional histogram and the radially binned data, the axes are calibrated assuming an excess noise $ \nex = 39\pm 6$, obtained from the change in variance between thermal and post-selected samples (see Methods). An independent quantum calibration of the gain and excess noise in the detection chain uses sideband asymmetry to determine the phonon temperature, and leads to values of gain and excess noise that are within roughly a factor of two. However, due to the sensitive dependence of system parameters such as the output efficiency on optical pump power, we found that other calibration approaches, specially when they require changing optical power, are unreliable (see Methods). The excess noise is dominated by heating after the optical pulse, an added noise of $ n_\text{m} = 2.4\pm 0.4 $ from the TWPA, and the total measurement efficiency of $ \sim 8\% $. Heating after the optical pulse adds extra thermal noise in the output microwave photon from the transducer. We estimate an added thermal noise $n_\text{n} = 1.6 \pm 0.5 $ in the propagating microwave photon with two different methods.  First, we use the excess noise from the heralding experiment and microwave readout efficiencies to infer the added thermal noise. Alternatively, we calculate the overlap between the independently measured time-domain heating after the optical pulse and the temporal mode of the microwave field to obtain the added noise (see Methods).

We have demonstrated the direct measurement of optically heralded microwave added photon states.  The heralding rate in our experiment is $ \sim \SI{15}{\hertz}$, limited by the total optical readout system efficiency of $\etasys = 1\%  $. While the heralding rate is comparable to other quantum systems for entanglement generation~\cite{hucul2015modular, humphreys2018deterministic,levine2018high, riedinger2018remote, van2022entangling}, we can drastically improve it by increasing the fiber-to-chip coupling efficiency~\cite{groblacher2013highly} and reducing optical filter insertion loss. These significantly improved heralding rates ($\sim \SI{}{\kilo\hertz} $) would be comparable to state-of-the-art coherence time of microwave~\cite{reagor2016quantum,romanenko2020three} and acoustic~\cite{maccabe2020nano} resonators. Another challenge is reducing the effects of induced thermal noise. Recent designs have emerged that demonstrate much better thermalization~\cite{ren2020two} allowing lower initial thermal occupation and added noise $ \nth, n_\text{n} \lesssim 0.1 $. Adopting these two-dimensional structures would also increase the achievable rates. The piezoelectric coupling rate of our transducer is also limited by the multi-mode nature of our microwave resonator. Moving to a single-mode microwave system~\cite{mirhosseini2020superconducting} would increase our phonon-to-microwave photon output efficiency from $\etamu \sim 35\%$ to close to $100\%$. To entangle two distant microwave systems, we will need to implement two copies of our transducer to produce indistinguishable optical photons inside two separate fridges. Overcoming frequency variations in different devices by frequency shifting in the optical domain~\cite{riedinger2018remote, levonian2022optical} or entanglement swapping with entangled optical photons at different frequencies from optical spontaneous parametric down conversion~\cite{krastanov2021optically} would relax the device frequency matching requirements. Our results show that with these improvements, piezo-optomechanical quantum frequency transducers that entangle distant quantum microwave systems are within reach.

\bibliographystyle{naturemag}

\bibliography{ref}

\begin{thebibliography}{10}
\expandafter\ifx\csname url\endcsname\relax
  \def\url#1{\texttt{#1}}\fi
\expandafter\ifx\csname urlprefix\endcsname\relax\def\urlprefix{URL }\fi
\providecommand{\bibinfo}[2]{#2}
\providecommand{\eprint}[2][]{\url{#2}}

\bibitem{kimble2008quantum}
\bibinfo{author}{Kimble, H.~J.}
\newblock \bibinfo{title}{The quantum internet}.
\newblock \emph{\bibinfo{journal}{Nature}} \textbf{\bibinfo{volume}{453}},
  \bibinfo{pages}{1023--1030} (\bibinfo{year}{2008}).

\bibitem{altman2021quantum}
\bibinfo{author}{Altman, E.} \emph{et~al.}
\newblock \bibinfo{title}{Quantum simulators: Architectures and opportunities}.
\newblock \emph{\bibinfo{journal}{PRX Quantum}} \textbf{\bibinfo{volume}{2}},
  \bibinfo{pages}{017003} (\bibinfo{year}{2021}).

\bibitem{alexeev2021quantum}
\bibinfo{author}{Alexeev, Y.} \emph{et~al.}
\newblock \bibinfo{title}{Quantum computer systems for scientific discovery}.
\newblock \emph{\bibinfo{journal}{PRX Quantum}} \textbf{\bibinfo{volume}{2}},
  \bibinfo{pages}{017001} (\bibinfo{year}{2021}).

\bibitem{awschalom2021development}
\bibinfo{author}{Awschalom, D.} \emph{et~al.}
\newblock \bibinfo{title}{Development of quantum interconnects (quics) for
  next-generation information technologies}.
\newblock \emph{\bibinfo{journal}{PRX Quantum}} \textbf{\bibinfo{volume}{2}},
  \bibinfo{pages}{017002} (\bibinfo{year}{2021}).

\bibitem{chen2021twin}
\bibinfo{author}{Chen, J.-P.} \emph{et~al.}
\newblock \bibinfo{title}{Twin-field quantum key distribution over a 511 km
  optical fibre linking two distant metropolitan areas}.
\newblock \emph{\bibinfo{journal}{Nature Photonics}}
  \textbf{\bibinfo{volume}{15}}, \bibinfo{pages}{570--575}
  (\bibinfo{year}{2021}).

\bibitem{devoret2013superconducting}
\bibinfo{author}{Devoret, M.~H.} \& \bibinfo{author}{Schoelkopf, R.~J.}
\newblock \bibinfo{title}{Superconducting circuits for quantum information: an
  outlook}.
\newblock \emph{\bibinfo{journal}{Science}} \textbf{\bibinfo{volume}{339}},
  \bibinfo{pages}{1169--1174} (\bibinfo{year}{2013}).

\bibitem{arute2019quantum}
\bibinfo{author}{Arute, F.} \emph{et~al.}
\newblock \bibinfo{title}{Quantum supremacy using a programmable
  superconducting processor}.
\newblock \emph{\bibinfo{journal}{Nature}} \textbf{\bibinfo{volume}{574}},
  \bibinfo{pages}{505--510} (\bibinfo{year}{2019}).

\bibitem{wu2021strong}
\bibinfo{author}{Wu, Y.} \emph{et~al.}
\newblock \bibinfo{title}{Strong quantum computational advantage using a
  superconducting quantum processor}.
\newblock \emph{\bibinfo{journal}{Physical Review Letters}}
  \textbf{\bibinfo{volume}{127}}, \bibinfo{pages}{180501}
  (\bibinfo{year}{2021}).

\bibitem{jurcevic2021demonstration}
\bibinfo{author}{Jurcevic, P.} \emph{et~al.}
\newblock \bibinfo{title}{Demonstration of quantum volume 64 on a
  superconducting quantum computing system}.
\newblock \emph{\bibinfo{journal}{Quantum Science and Technology}}
  \textbf{\bibinfo{volume}{6}}, \bibinfo{pages}{025020} (\bibinfo{year}{2021}).

\bibitem{mirhosseini2020superconducting}
\bibinfo{author}{Mirhosseini, M.}, \bibinfo{author}{Sipahigil, A.},
  \bibinfo{author}{Kalaee, M.} \& \bibinfo{author}{Painter, O.}
\newblock \bibinfo{title}{Superconducting qubit to optical photon
  transduction}.
\newblock \emph{\bibinfo{journal}{Nature}} \textbf{\bibinfo{volume}{588}},
  \bibinfo{pages}{599--603} (\bibinfo{year}{2020}).

\bibitem{krastanov2021optically}
\bibinfo{author}{Krastanov, S.} \emph{et~al.}
\newblock \bibinfo{title}{Optically heralded entanglement of superconducting
  systems in quantum networks}.
\newblock \emph{\bibinfo{journal}{Physical Review Letters}}
  \textbf{\bibinfo{volume}{127}}, \bibinfo{pages}{040503}
  (\bibinfo{year}{2021}).

\bibitem{pompili2021realization}
\bibinfo{author}{Pompili, M.} \emph{et~al.}
\newblock \bibinfo{title}{Realization of a multinode quantum network of remote
  solid-state qubits}.
\newblock \emph{\bibinfo{journal}{Science}} \textbf{\bibinfo{volume}{372}},
  \bibinfo{pages}{259--264} (\bibinfo{year}{2021}).

\bibitem{daiss2021quantum}
\bibinfo{author}{Daiss, S.} \emph{et~al.}
\newblock \bibinfo{title}{A quantum-logic gate between distant quantum-network
  modules}.
\newblock \emph{\bibinfo{journal}{Science}} \textbf{\bibinfo{volume}{371}},
  \bibinfo{pages}{614--617} (\bibinfo{year}{2021}).

\bibitem{ofek2016extending}
\bibinfo{author}{Ofek, N.} \emph{et~al.}
\newblock \bibinfo{title}{Extending the lifetime of a quantum bit with error
  correction in superconducting circuits}.
\newblock \emph{\bibinfo{journal}{Nature}} \textbf{\bibinfo{volume}{536}},
  \bibinfo{pages}{441--445} (\bibinfo{year}{2016}).

\bibitem{krinner2022realizing}
\bibinfo{author}{Krinner, S.} \emph{et~al.}
\newblock \bibinfo{title}{Realizing repeated quantum error correction in a
  distance-three surface code}.
\newblock \emph{\bibinfo{journal}{Nature}} \textbf{\bibinfo{volume}{605}},
  \bibinfo{pages}{669--674} (\bibinfo{year}{2022}).

\bibitem{serafini2006distributed}
\bibinfo{author}{Serafini, A.}, \bibinfo{author}{Mancini, S.} \&
  \bibinfo{author}{Bose, S.}
\newblock \bibinfo{title}{Distributed quantum computation via optical fibers}.
\newblock \emph{\bibinfo{journal}{Physical Review Letters}}
  \textbf{\bibinfo{volume}{96}}, \bibinfo{pages}{010503}
  (\bibinfo{year}{2006}).

\bibitem{gottesman2012longer}
\bibinfo{author}{Gottesman, D.}, \bibinfo{author}{Jennewein, T.} \&
  \bibinfo{author}{Croke, S.}
\newblock \bibinfo{title}{Longer-baseline telescopes using quantum repeaters}.
\newblock \emph{\bibinfo{journal}{Physical Review Letters}}
  \textbf{\bibinfo{volume}{109}}, \bibinfo{pages}{070503}
  (\bibinfo{year}{2012}).

\bibitem{brady2022entangled}
\bibinfo{author}{Brady, A.~J.} \emph{et~al.}
\newblock \bibinfo{title}{Entangled {Sensor-Networks for Dark-Matter
  Searches}}.
\newblock \emph{\bibinfo{journal}{PRX Quantum}} \textbf{\bibinfo{volume}{3}},
  \bibinfo{pages}{030333} (\bibinfo{year}{2022}).

\bibitem{shor2000simple}
\bibinfo{author}{Shor, P.~W.} \& \bibinfo{author}{Preskill, J.}
\newblock \bibinfo{title}{Simple proof of security of the {BB84} quantum key
  distribution protocol}.
\newblock \emph{\bibinfo{journal}{Physical Review Letters}}
  \textbf{\bibinfo{volume}{85}}, \bibinfo{pages}{441} (\bibinfo{year}{2000}).

\bibitem{han2021microwave}
\bibinfo{author}{Han, X.}, \bibinfo{author}{Fu, W.}, \bibinfo{author}{Zou,
  C.-L.}, \bibinfo{author}{Jiang, L.} \& \bibinfo{author}{Tang, H.~X.}
\newblock \bibinfo{title}{Microwave-optical quantum frequency conversion}.
\newblock \emph{\bibinfo{journal}{Optica}} \textbf{\bibinfo{volume}{8}},
  \bibinfo{pages}{1050--1064} (\bibinfo{year}{2021}).

\bibitem{aspelmeyer2014cavity}
\bibinfo{author}{Aspelmeyer, M.}, \bibinfo{author}{Kippenberg, T.~J.} \&
  \bibinfo{author}{Marquardt, F.}
\newblock \bibinfo{title}{Cavity optomechanics}.
\newblock \emph{\bibinfo{journal}{Reviews of Modern Physics}}
  \textbf{\bibinfo{volume}{86}}, \bibinfo{pages}{1391} (\bibinfo{year}{2014}).

\bibitem{teufel2011circuit}
\bibinfo{author}{Teufel, J.~D.} \emph{et~al.}
\newblock \bibinfo{title}{Circuit cavity electromechanics in the
  strong-coupling regime}.
\newblock \emph{\bibinfo{journal}{Nature}} \textbf{\bibinfo{volume}{471}},
  \bibinfo{pages}{204--208} (\bibinfo{year}{2011}).

\bibitem{andrews2014bidirectional}
\bibinfo{author}{Andrews, R.~W.} \emph{et~al.}
\newblock \bibinfo{title}{Bidirectional and efficient conversion between
  microwave and optical light}.
\newblock \emph{\bibinfo{journal}{Nature Physics}}
  \textbf{\bibinfo{volume}{10}}, \bibinfo{pages}{321--326}
  (\bibinfo{year}{2014}).

\bibitem{o2010quantum}
\bibinfo{author}{O’Connell, A.~D.} \emph{et~al.}
\newblock \bibinfo{title}{Quantum ground state and single-phonon control of a
  mechanical resonator}.
\newblock \emph{\bibinfo{journal}{Nature}} \textbf{\bibinfo{volume}{464}},
  \bibinfo{pages}{697--703} (\bibinfo{year}{2010}).

\bibitem{bochmann2013nanomechanical}
\bibinfo{author}{Bochmann, J.}, \bibinfo{author}{Vainsencher, A.},
  \bibinfo{author}{Awschalom, D.~D.} \& \bibinfo{author}{Cleland, A.~N.}
\newblock \bibinfo{title}{Nanomechanical coupling between microwave and optical
  photons}.
\newblock \emph{\bibinfo{journal}{Nature Physics}}
  \textbf{\bibinfo{volume}{9}}, \bibinfo{pages}{712--716}
  (\bibinfo{year}{2013}).

\bibitem{zeuthen2020figures}
\bibinfo{author}{Zeuthen, E.}, \bibinfo{author}{Schliesser, A.},
  \bibinfo{author}{S{\o}rensen, A.~S.} \& \bibinfo{author}{Taylor, J.~M.}
\newblock \bibinfo{title}{Figures of merit for quantum transducers}.
\newblock \emph{\bibinfo{journal}{Quantum Science and Technology}}
  \textbf{\bibinfo{volume}{5}}, \bibinfo{pages}{034009} (\bibinfo{year}{2020}).

\bibitem{han2020cavity}
\bibinfo{author}{Han, X.} \emph{et~al.}
\newblock \bibinfo{title}{Cavity piezo-mechanics for
  superconducting-nanophotonic quantum interface}.
\newblock \emph{\bibinfo{journal}{Nature Communications}}
  \textbf{\bibinfo{volume}{11}}, \bibinfo{pages}{1--8} (\bibinfo{year}{2020}).

\bibitem{brubaker2022optomechanical}
\bibinfo{author}{Brubaker, B.~M.} \emph{et~al.}
\newblock \bibinfo{title}{Optomechanical ground-state cooling in a continuous
  and efficient electro-optic transducer}.
\newblock \emph{\bibinfo{journal}{Physical Review X}}
  \textbf{\bibinfo{volume}{12}}, \bibinfo{pages}{021062}
  (\bibinfo{year}{2022}).

\bibitem{delaney2022superconducting}
\bibinfo{author}{Delaney, R.} \emph{et~al.}
\newblock \bibinfo{title}{Superconducting-qubit readout via low-backaction
  electro-optic transduction}.
\newblock \emph{\bibinfo{journal}{Nature}} \textbf{\bibinfo{volume}{606}},
  \bibinfo{pages}{489--493} (\bibinfo{year}{2022}).

\bibitem{pop2017laterally}
\bibinfo{author}{Pop, F.~V.}, \bibinfo{author}{Kochhar, A.~S.},
  \bibinfo{author}{Vidal-Alvarez, G.} \& \bibinfo{author}{Piazza, G.}
\newblock \bibinfo{title}{Laterally vibrating lithium niobate {MEMS} resonators
  with 30\% electromechanical coupling coefficient}.
\newblock In \emph{\bibinfo{booktitle}{2017 IEEE 30th International Conference
  on Micro Electro Mechanical Systems (MEMS)}}, \bibinfo{pages}{966--969}
  (\bibinfo{organization}{IEEE}, \bibinfo{year}{2017}).

\bibitem{safavi2019controlling}
\bibinfo{author}{Safavi-Naeini, A.~H.}, \bibinfo{author}{Van~Thourhout, D.},
  \bibinfo{author}{Baets, R.} \& \bibinfo{author}{Van~Laer, R.}
\newblock \bibinfo{title}{Controlling phonons and photons at the wavelength
  scale: integrated photonics meets integrated phononics}.
\newblock \emph{\bibinfo{journal}{Optica}} \textbf{\bibinfo{volume}{6}},
  \bibinfo{pages}{213--232} (\bibinfo{year}{2019}).

\bibitem{chan2012optimized}
\bibinfo{author}{Chan, J.}, \bibinfo{author}{Safavi-Naeini, A.~H.},
  \bibinfo{author}{Hill, J.~T.}, \bibinfo{author}{Meenehan, S.} \&
  \bibinfo{author}{Painter, O.}
\newblock \bibinfo{title}{Optimized optomechanical crystal cavity with acoustic
  radiation shield}.
\newblock \emph{\bibinfo{journal}{Applied Physics Letters}}
  \textbf{\bibinfo{volume}{101}}, \bibinfo{pages}{081115}
  (\bibinfo{year}{2012}).

\bibitem{xu2019frequency}
\bibinfo{author}{Xu, M.}, \bibinfo{author}{Han, X.}, \bibinfo{author}{Fu, W.},
  \bibinfo{author}{Zou, C.-L.} \& \bibinfo{author}{Tang, H.~X.}
\newblock \bibinfo{title}{Frequency-tunable high-{Q} superconducting resonators
  via wireless control of nonlinear kinetic inductance}.
\newblock \emph{\bibinfo{journal}{Applied Physics Letters}}
  \textbf{\bibinfo{volume}{114}}, \bibinfo{pages}{192601}
  (\bibinfo{year}{2019}).

\bibitem{macklin2015near}
\bibinfo{author}{Macklin, C.} \emph{et~al.}
\newblock \bibinfo{title}{A near--quantum-limited {Josephson} traveling-wave
  parametric amplifier}.
\newblock \emph{\bibinfo{journal}{Science}} \textbf{\bibinfo{volume}{350}},
  \bibinfo{pages}{307--310} (\bibinfo{year}{2015}).

\bibitem{eichler2011experimental}
\bibinfo{author}{Eichler, C.} \emph{et~al.}
\newblock \bibinfo{title}{Experimental state tomography of itinerant single
  microwave photons}.
\newblock \emph{\bibinfo{journal}{Physical Review Letters}}
  \textbf{\bibinfo{volume}{106}}, \bibinfo{pages}{220503}
  (\bibinfo{year}{2011}).

\bibitem{patel2021room}
\bibinfo{author}{Patel, R.~N.} \emph{et~al.}
\newblock \bibinfo{title}{Room-temperature mechanical resonator with a single
  added or subtracted phonon}.
\newblock \emph{\bibinfo{journal}{Physical Review Letters}}
  \textbf{\bibinfo{volume}{127}}, \bibinfo{pages}{133602}
  (\bibinfo{year}{2021}).

\bibitem{hucul2015modular}
\bibinfo{author}{Hucul, D.} \emph{et~al.}
\newblock \bibinfo{title}{Modular entanglement of atomic qubits using photons
  and phonons}.
\newblock \emph{\bibinfo{journal}{Nature Physics}}
  \textbf{\bibinfo{volume}{11}}, \bibinfo{pages}{37--42}
  (\bibinfo{year}{2015}).

\bibitem{humphreys2018deterministic}
\bibinfo{author}{Humphreys, P.~C.} \emph{et~al.}
\newblock \bibinfo{title}{Deterministic delivery of remote entanglement on a
  quantum network}.
\newblock \emph{\bibinfo{journal}{Nature}} \textbf{\bibinfo{volume}{558}},
  \bibinfo{pages}{268--273} (\bibinfo{year}{2018}).

\bibitem{levine2018high}
\bibinfo{author}{Levine, H.} \emph{et~al.}
\newblock \bibinfo{title}{High-fidelity control and entanglement of
  {Rydberg-atom} qubits}.
\newblock \emph{\bibinfo{journal}{Physical Review Letters}}
  \textbf{\bibinfo{volume}{121}}, \bibinfo{pages}{123603}
  (\bibinfo{year}{2018}).

\bibitem{riedinger2018remote}
\bibinfo{author}{Riedinger, R.} \emph{et~al.}
\newblock \bibinfo{title}{Remote quantum entanglement between two
  micromechanical oscillators}.
\newblock \emph{\bibinfo{journal}{Nature}} \textbf{\bibinfo{volume}{556}},
  \bibinfo{pages}{473--477} (\bibinfo{year}{2018}).

\bibitem{van2022entangling}
\bibinfo{author}{van Leent, T.} \emph{et~al.}
\newblock \bibinfo{title}{Entangling single atoms over 33 km telecom fibre}.
\newblock \emph{\bibinfo{journal}{Nature}} \textbf{\bibinfo{volume}{607}},
  \bibinfo{pages}{69--73} (\bibinfo{year}{2022}).

\bibitem{groblacher2013highly}
\bibinfo{author}{Gr{\"o}blacher, S.}, \bibinfo{author}{Hill, J.~T.},
  \bibinfo{author}{Safavi-Naeini, A.~H.}, \bibinfo{author}{Chan, J.} \&
  \bibinfo{author}{Painter, O.}
\newblock \bibinfo{title}{Highly efficient coupling from an optical fiber to a
  nanoscale silicon optomechanical cavity}.
\newblock \emph{\bibinfo{journal}{Applied Physics Letters}}
  \textbf{\bibinfo{volume}{103}}, \bibinfo{pages}{181104}
  (\bibinfo{year}{2013}).

\bibitem{reagor2016quantum}
\bibinfo{author}{Reagor, M.} \emph{et~al.}
\newblock \bibinfo{title}{Quantum memory with millisecond coherence in circuit
  {QED}}.
\newblock \emph{\bibinfo{journal}{Physical Review B}}
  \textbf{\bibinfo{volume}{94}}, \bibinfo{pages}{014506}
  (\bibinfo{year}{2016}).

\bibitem{romanenko2020three}
\bibinfo{author}{Romanenko, A.} \emph{et~al.}
\newblock \bibinfo{title}{Three-dimensional superconducting resonators at {T$<$
  20 mK} with photon lifetimes up to $\tau$= 2 s}.
\newblock \emph{\bibinfo{journal}{Physical Review Applied}}
  \textbf{\bibinfo{volume}{13}}, \bibinfo{pages}{034032}
  (\bibinfo{year}{2020}).

\bibitem{maccabe2020nano}
\bibinfo{author}{MacCabe, G.~S.} \emph{et~al.}
\newblock \bibinfo{title}{Nano-acoustic resonator with ultralong phonon
  lifetime}.
\newblock \emph{\bibinfo{journal}{Science}} \textbf{\bibinfo{volume}{370}},
  \bibinfo{pages}{840--843} (\bibinfo{year}{2020}).

\bibitem{ren2020two}
\bibinfo{author}{Ren, H.} \emph{et~al.}
\newblock \bibinfo{title}{Two-dimensional optomechanical crystal cavity with
  high quantum cooperativity}.
\newblock \emph{\bibinfo{journal}{Nature Communications}}
  \textbf{\bibinfo{volume}{11}}, \bibinfo{pages}{1--10} (\bibinfo{year}{2020}).

\bibitem{levonian2022optical}
\bibinfo{author}{Levonian, D.} \emph{et~al.}
\newblock \bibinfo{title}{Optical entanglement of distinguishable quantum
  emitters}.
\newblock \emph{\bibinfo{journal}{Physical Review Letters}}
  \textbf{\bibinfo{volume}{128}}, \bibinfo{pages}{213602}
  (\bibinfo{year}{2022}).

\bibitem{fiaschi2021optomechanical}
\bibinfo{author}{Fiaschi, N.} \emph{et~al.}
\newblock \bibinfo{title}{Optomechanical quantum teleportation}.
\newblock \emph{\bibinfo{journal}{Nature Photonics}}
  \textbf{\bibinfo{volume}{15}}, \bibinfo{pages}{817--821}
  (\bibinfo{year}{2021}).

\bibitem{peairs2020continuous}
\bibinfo{author}{Peairs, G.} \emph{et~al.}
\newblock \bibinfo{title}{Continuous and time-domain coherent signal conversion
  between optical and microwave frequencies}.
\newblock \emph{\bibinfo{journal}{Physical Review Applied}}
  \textbf{\bibinfo{volume}{14}}, \bibinfo{pages}{061001}
  (\bibinfo{year}{2020}).

\bibitem{arrangoiz2019resolving}
\bibinfo{author}{Arrangoiz-Arriola, P.} \emph{et~al.}
\newblock \bibinfo{title}{Resolving the energy levels of a nanomechanical
  oscillator}.
\newblock \emph{\bibinfo{journal}{Nature}} \textbf{\bibinfo{volume}{571}},
  \bibinfo{pages}{537--540} (\bibinfo{year}{2019}).

\bibitem{meitl2006transfer}
\bibinfo{author}{Meitl, M.~A.} \emph{et~al.}
\newblock \bibinfo{title}{Transfer printing by kinetic control of adhesion to
  an elastomeric stamp}.
\newblock \emph{\bibinfo{journal}{Nature Materials}}
  \textbf{\bibinfo{volume}{5}}, \bibinfo{pages}{33--38} (\bibinfo{year}{2006}).

\bibitem{jiang2019lithium}
\bibinfo{author}{Jiang, W.} \emph{et~al.}
\newblock \bibinfo{title}{Lithium niobate piezo-optomechanical crystals}.
\newblock \emph{\bibinfo{journal}{Optica}} \textbf{\bibinfo{volume}{6}},
  \bibinfo{pages}{845--853} (\bibinfo{year}{2019}).

\bibitem{Wenner2011}
\bibinfo{author}{Wenner, J.} \emph{et~al.}
\newblock \bibinfo{title}{Wirebond crosstalk and cavity modes in large chip
  mounts for superconducting qubits}.
\newblock \emph{\bibinfo{journal}{Supercond. Sci. Technol}}
  \textbf{\bibinfo{volume}{24}}, \bibinfo{pages}{65001--65008}
  (\bibinfo{year}{2011}).

\bibitem{huang2021microwave}
\bibinfo{author}{Huang, S.} \emph{et~al.}
\newblock \bibinfo{title}{Microwave package design for superconducting quantum
  processors}.
\newblock \emph{\bibinfo{journal}{PRX Quantum}} \textbf{\bibinfo{volume}{2}},
  \bibinfo{pages}{020306} (\bibinfo{year}{2021}).

\bibitem{zhong2021deterministic}
\bibinfo{author}{Zhong, Y.} \emph{et~al.}
\newblock \bibinfo{title}{Deterministic multi-qubit entanglement in a quantum
  network}.
\newblock \emph{\bibinfo{journal}{Nature}} \textbf{\bibinfo{volume}{590}},
  \bibinfo{pages}{571--575} (\bibinfo{year}{2021}).

\end{thebibliography}

\section*{Methods}

\renewcommand{\figurename}{}
\renewcommand{\thefigure}{Extended Data Fig.~\arabic{figure}}
\setcounter{figure}{0}

\renewcommand{\tablename}{Extended Data Table}
\setcounter{table}{0}

\subsection*{Device parameters}

Extended Data Table~\ref{tab:parameters} shows the device parameters for the optical, mechanical, microwave modes and the coupling rates. The measurement method used to obtain each parameter is also listed. We observe that the mechanical frequency and intrinsic loss rate are different under different optical powers. When the optical pump is turned off, $\gammai$ is reduced by a factor of $\sim 3$ and the mechanical frequency red-shifts by $\sim \SI{700}{\kilo\hertz}$. Variation of the mechanical frequency and intrinsic loss rate are likely due to a combination of optically induced heating and saturation of two-level systems. Reducing the frequency fluctuation and improving the intrinsic mechanical quality factor will be the subject of future effort.

\begin{table*}
\caption{\label{tab:parameters}\textbf{Device parameters.}}
\begin{center}
\begin{tabular}{c  c   c }
\hline
Parameter & Value & method\\
\hline
$\omega_\text{o}/2\pi$ & \SI{193.53}{\tera\hertz} & Laser wavelength sweep\\
$\kappa_\text{o}/2\pi$ & \SI{1.122}{\giga\hertz} & EIT ($ S_\text{oo}$)\\
$ \kappa_\text{o,e}/2\pi $ & \SI{0.561}{\giga\hertz} & EIT\\
$g_\text{o}/2\pi$ & \SI{413}{\kilo\hertz} & EIT\\
$\omega_\text{m}/2\pi$ & $\sim$ \SI{3.596}{\giga\hertz} & Microwave $S_{\upmu\upmu}$\\
$\gammai/2\pi$ (pump off) & \SI{0.36}{\mega\hertz} & Microwave $S_{\upmu\upmu}$\\
$\gammai/2\pi$ (pump on) & \SI{1.07}{\mega\hertz} & Microwave $S_{\upmu\upmu}$\\
$\omega_{\upmu}/2\pi$ & \SI{3.5958}{\giga\hertz} & Microwave $S_{\upmu\upmu}$\\
$\kappa_{\upmu}/2\pi$ & \SI{3.06}{\mega\hertz} & Microwave $S_{\upmu\upmu}$\\
$\kappa_{\upmu,\text{e}}/2\pi$ & \SI{3.04}{\mega\hertz} & Microwave $S_{\upmu\upmu}$\\
$g_{\upmu}/2\pi$ & \SI{424}{\kilo\hertz} & Microwave $S_{\upmu\upmu}$\\
\hline
\end{tabular}
\end{center}
\end{table*}

\subsection*{Tunable high-impedance waveguide}

\begin{figure*}[tb]
\centering
\includegraphics[scale=1.]{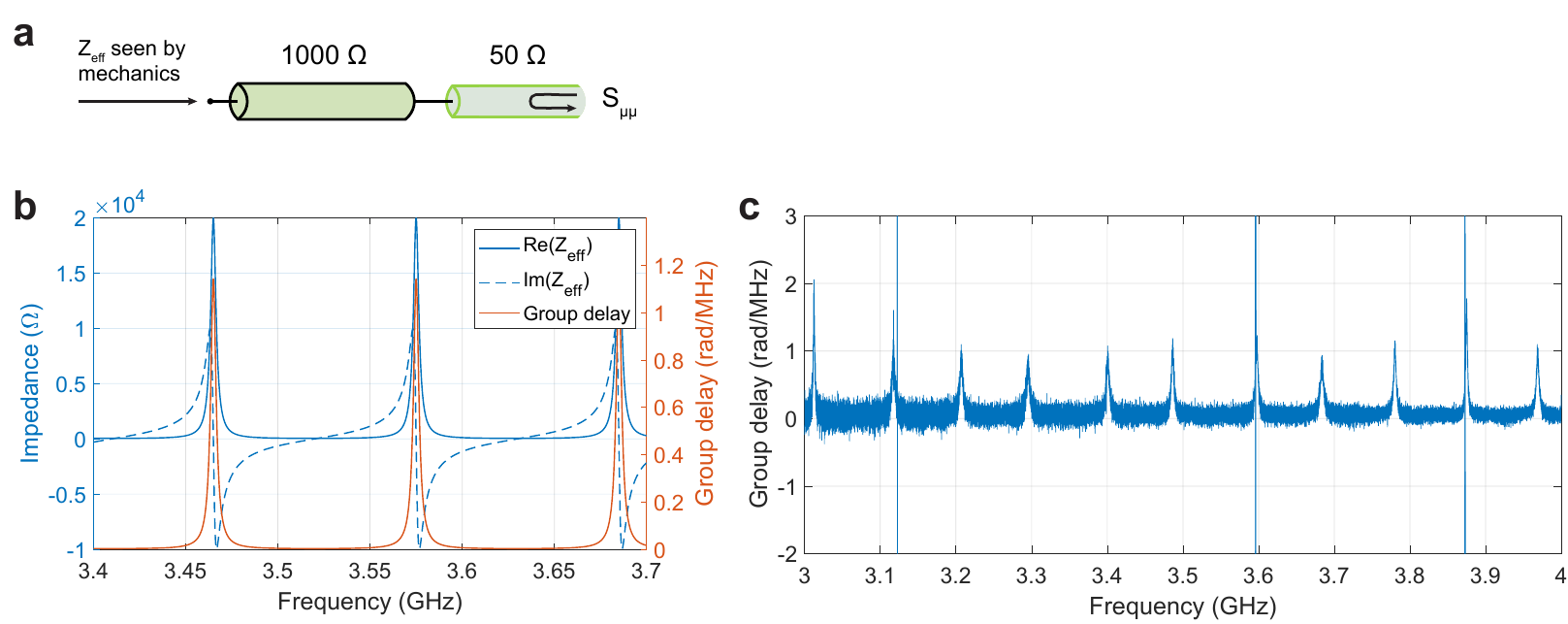}
\caption{\label{SI-Fig:microwave} \textbf{Microwave high-impedance waveguide characterization.} \textbf{a}, Schematics of the high impedance (high-Z) waveguide. The mechanical external coupling is approximately proportional to the impedance of the environment. Thus the effective impedance $\Zeff$ looking from the mechanics side into the high-Z waveguide gives us intuition on the enhanced external coupling. \textbf{b}, Calculated $\Zeff$ and the group delay with a free spectral range of $ \SI{110}{\mega\hertz} $ and a waveguide characteristic impedance $Z = \SI{1000}{\ohm} $. \textbf{c}, Measured group delay of the high-Z waveguide. The coupled mechanical modes appear as sharp peaks in group delay.}
\end{figure*}

\begin{figure*}[tb]
\centering
\includegraphics[scale=0.55]{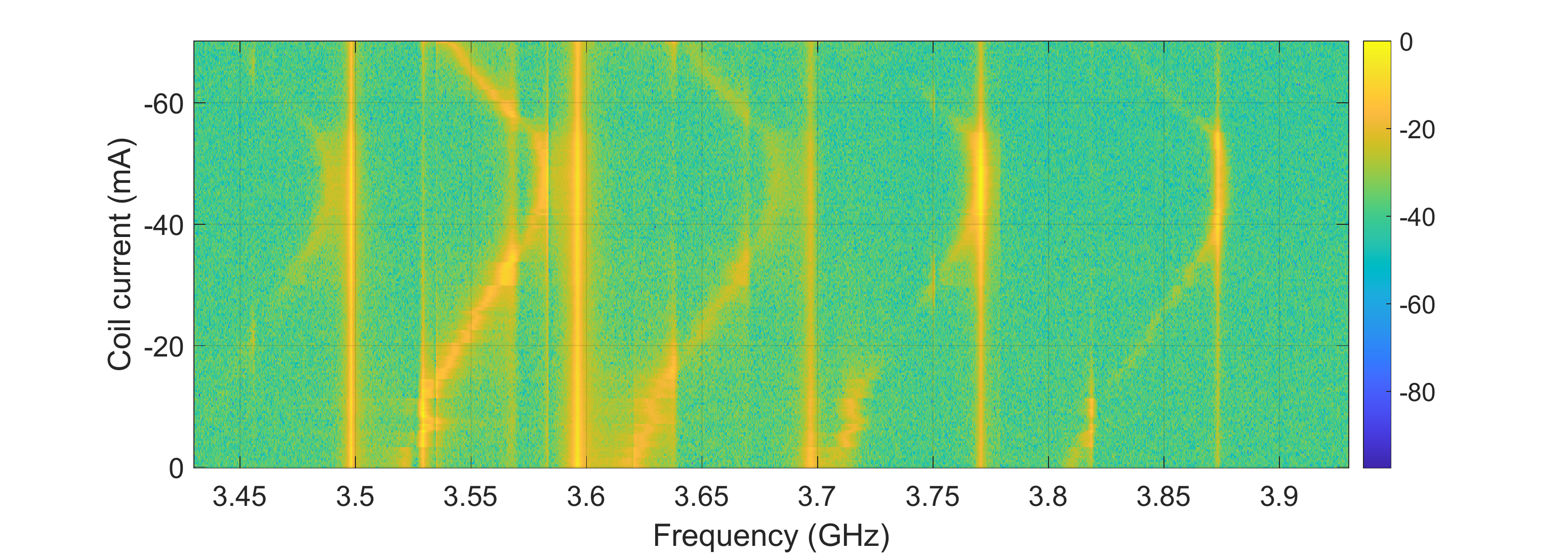}
\caption{\label{SI-Fig:coil-sweep} \textbf{Microwave-to-optical conversion with different microwave resonator frequencies.} Plot of normalized microwave-to-optical conversion S parameter as a function of frequency and coil current.}
\end{figure*}

The high-impedance (high-Z) waveguide is realized with a hybrid aluminum-NbTiN coplanar waveguide (CPW). The ground plane of the CPW is made of e-beam evaporated aluminum. The center conductor of the CPW is made of thin and narrow NbTiN nanowires to achieve high kinetic inductance. The thickness of the NbTiN layer is \SI{10}{\nano\meter}, deposited by StarCryo on high-resistivity silicon ($ \rho > \SI{10}{\kilo\ohm\per\centi\meter}$) substrate from WaferPro. The width of the nanowire is \SI{500}{\nano\meter}. Square-shaped loops with \SI{50}{\micro\meter} width formed by the nanowires along the center conductor of the CPW allow wireless tuning of the kinetic inductance through an external magnetic field~\cite{xu2019frequency}. The distance between the edges of the ground plane is chosen to be \SI{150}{\micro\meter} to reduce the capacitance. A home-made tuning coil with \SI{50}{\milli\meter} diameter and $\sim 2000$ turns of NbTi wire (Supercon, Inc., SC-T48B-M-0.10mm) generates $\sim\SI{0.01}{\milli\tesla\per\milli\ampere}$ at the chip. We observe negligible heating from the coil with up to \SI{200}{\milli\ampere} continuous current thanks to zero heat dissipation in the superconducting NbTi wire. The total length of the high-Z CPW is $\sim\SI{65}{\milli\meter}$, giving rise to standing wave resonances with a free spectral range (FSR) of $\sim\SI{110}{\mega\hertz}$ (\ref{SI-Fig:microwave}). The waveguide resonances can be frequency-tuned by more than \SI{150}{\mega\hertz}, larger than the FSR of the resonances, allowing us to match a waveguide resonance to a mechanical mode over a broad frequency range.

We show the measured microwave-to-optical conversion S parameter in \ref{SI-Fig:coil-sweep} as we vary the coil current. Mechanical resonances of the transducer are not affected by the coil current, while the conversion is enhanced when the microwave modes are resonant with the mechanical modes. The microwave modes reach their maximal frequencies at non-zero current due to non-zero trapped external flux in the tuning loops from a non-zero background magnetic field during the cooldown.

\subsection*{Thermal occupation measurement and time-domain heating from the optical pulse}
\label{subs:thermal}

\begin{figure*}[tb]
\centering
\includegraphics[scale=1]{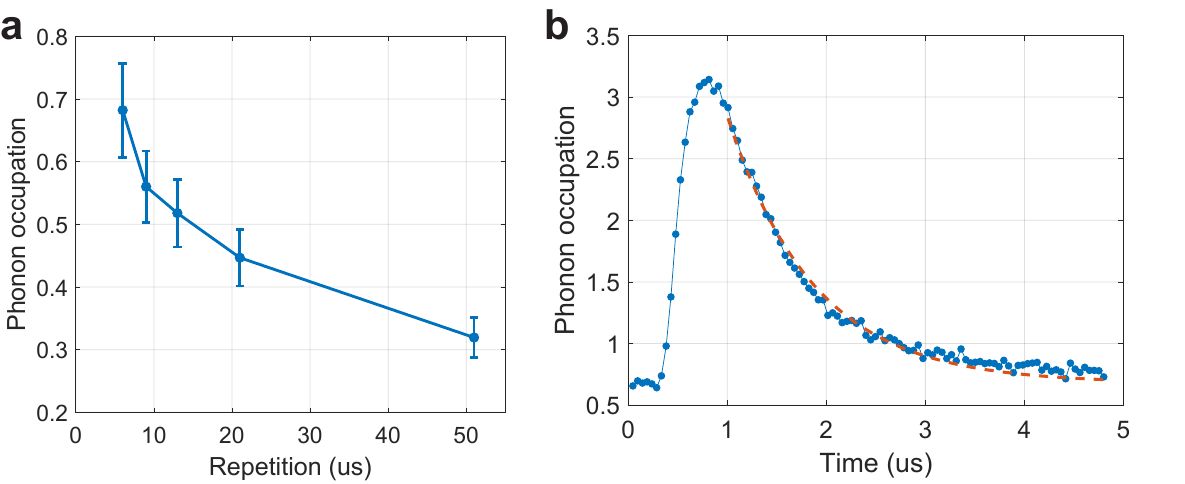}
\caption{\label{SI-Fig:pulsed-char} \textbf{Thermal occupation and heating.} \textbf{a}, Thermal phonon occupation of the mechanical mode versus repetition time of the optical pump pulse. As explained in Methods, the thermal occupation is calculated from sideband asymmetry between the blue and red detuned pump pulse. \textbf{b}, Time-domain heating of the optical pulse at \SI{6}{\micro\second} repetition, measured with microwave readout of the mechanical mode. Variance of the microwave noise is measured and calibrated to thermal phonon number with optical sideband asymmetry. Dashed red line is the exponential fit of the decaying tail of the thermal phonon occupation.}
\end{figure*}

Thermal occupation of the mechanical mode $\nth $ is measured with sideband asymmetry by comparing the single-photon detection rates of the optomechanically scattered optical sidebands. Two tunable external cavity diode lasers (Pure Photonics PPCL300) are locked to detunings $\Delta_\pm \equiv \omega_{\pm} - \omegao = \pm \omegam$ with respect to the optical mode $\omegao$ as blue and red detuned pumps. Due to imperfect laser detunings and the narrow linewidths of the optical filter cavities, the count rates are recorded as the filter cavities are tuned by $\sim\SI{20}{\mega\hertz}$ between the blue-detuned and red-detuned sideband count rate measurement to optimize the sideband count rate. The insertion loss between counts at the two frequencies could be different. To calibrate the varying insertion loss, we drive a coherent phonon occupation $\ncoh$ with a microwave pulse and measure its sideband count rate in addition to sideband count rate from the thermal phonons $\nth$ in an interleaved fashion. Ratio of the sideband count rates between $\nth$ and $\ncoh$ is independent of the filter insertion loss,
\begin{equation}
    R_\text{r} = \frac{\ncoh}{\nth},~ R_\text{b} = \frac{\ncoh + 1}{\nth + 1},
\end{equation}
where the subscript denotes the laser detuning. By further comparing these ratios between blue and red detuned pump, we find
\begin{equation}
    \frac{R_\text{b} - 1}{R_\text{r} - 1} = \frac{\nth}{\nth + 1} \equiv A,
\end{equation}
giving us the sideband asymmetry ratio $A$ and thermal phonon number $ \nth = 1/(1/A-1) $. \ref{SI-Fig:pulsed-char}a shows the measured thermal occupation versus different repetition rates of the optical pulse. We find that our repetition rate and thermal occupation are comparable to similar optomechanical systems~\cite{fiaschi2021optomechanical}.

The microwave port of the transducer allow us to monitor the microwave noise from the transducer, which is dominated by converted thermal mechanical noise when the transducer is under pulsed optical pump. As shown in \ref{SI-Fig:pulsed-char}b, we measured the temporal heating of the mechanical mode by using a series of consecutive demodulation windows, each of duration \SI{48}{\nano\second}. Changes in the variance of the demodulated data correspond to varying thermal noise in the microwave signal, which can be calibrated to a varying thermal phonon occupation by the initial thermal occupation measured with optical sideband asymmetry.

\subsection*{Conversion efficiency and added noise measurement}

The conversion efficiency of the transducer is defined as
\begin{equation}
\label{eqn:conversion-efficiency}
    \eta_{\upmu \text{o}} \equiv \frac{\dot N_{\text{out}, \upmu} }{ \dot N_{\text{in}, \text{o}} },~ \eta_{ \text{o} \upmu} \equiv \frac{\dot N_{\text{out}, \text{o}} }{ \dot N_{\text{in}, \upmu } },
\end{equation}
where $\dot N_{\text{in(out)}, \upmu (\text{o})}$ is the input (output) microwave (optical) photon flux.

The input and output optical photon flux are measured with the sideband filter and the single photon detector (SPD). For the output photon flux from microwave-to-optical conversion, it can be directly measured with the optical setup shown in \ref{SI-Fig:optical-setup}. The system detection efficiency of the optical detection setup, including the insertion loss of the optical switches, the isolator, the two sideband filters, and the SPD efficiency, is measured independently to be $\eta_\text{SPD} = 9.9\%$. The optical insertion loss within the dilution refrigerator, and the one-way coupling efficiency between the lensed fiber and the on-chip waveguide, is measured to be $\eta_\text{in-fridge} = 25.4\%$ in total. We measure the insertion loss of the output circulator to be $\eta_\text{circ} = 77\%$. Together we calculate an overall optical setup efficiency from on-chip photonic waveguide to the SPD to be $\eta_\text{setup} = 2\%$. Note that the optical mode external coupling efficiency $\etao = 50\%$ is not included in the setup efficiency, while it is included in the conversion efficiency $\eta$ and system efficiency $\etasys$. The measured count rate at the SPD and all the output insertion losses are used to calculate the output photon flux at the device.

For the input photon flux during the optical-to-microwave conversion, the dilution refrigerator is bypassed in the optical circuit, i.e., the output of the circulator that is originally connected to the fridge input port is connected to the optical switch after the circulator. As a result, the same output optical circuit is used to measure the sideband photon flux at the fridge input port. The in-fridge insertion loss $\eta_\text{in-fridge}$ is then used to calculate the input photon flux at the device.

The microwave photon flux at the device is not directly measurable, there is also no independent way of measuring the microwave input attenuation and the output amplification $\Gm$ separately in our experiment. However, $\Gm$ can be calculated using the measured optical photon flux and external microwave power assuming the conversion efficiencies are equal between the two directions. More specifically,
\begin{equation}
\label{eqn:MW-flux}
    \dot N_{\text{out}, \upmu} = \frac{P_{\upmu, \upmu \text{o}}}{\hbar \omega_\upmu}\frac{1}{\Gm},~ \dot N_{\text{in}, \upmu} = \frac{P_{\upmu,  \text{o}\upmu} }{\hbar \omega_\upmu}\frac{1}{\Gm |S_{\upmu \upmu}|^2},
\end{equation}
where $P_{\upmu, \upmu \text{o}(\text{o} \upmu ) }$ denotes the output microwave power measured at the real-time spectrum analyzer (RSA) for the optical-to-microwave (microwave-to-optical) conversion. $\Gm$ converts between the microwave photon flux at the RSA and the flux leaving the transducer $ \dot N_{\text{out}, \upmu} $. For the microwave input flux, the output flux is converted to input using microwave reflection $S_{\upmu \upmu} $ independently measured with the VNA. Substituting Eq.~\ref{eqn:MW-flux} in Eq.~\ref{eqn:conversion-efficiency}, we find that $ \eta_{\upmu \text{o}} = \eta_{\text{o}\upmu } $ leads to
\begin{eqnarray}
    \Gm &=& \left( \frac{P_{\upmu, \upmu \text{o}} P_{\upmu,  \text{o}\upmu}/(\hbar \omega_\upmu)^2 }{ \dot N_{\text{in}, \text{o}}   \dot N_{\text{out}, \text{o}}  |S_{\upmu \upmu}|^2 } \right)^{1/2},\\
    \eta_{\upmu \text{o}}&=& \eta_{ \text{o} \upmu} = \left( \frac{\dot N_{\text{out}, \text{o}} }{ \dot N_{\text{in}, \text{o}} } \frac{ P_{\upmu, \upmu \text{o}} |S_{\upmu \upmu}|^2 }{ P_{\upmu,  \text{o}\upmu} } \right)^{1/2}.
\end{eqnarray}

$\Gm$ can be used to calculate the added noise in the microwave output during the optical-to-microwave conversion, which is directly measured by the RSA. We find it to be $4.9$ at the microwave output of the device, and the corresponding added noise referred to the input is $n_{\text{added}, \upmu\text{o}} = 99 \pm 10$. Based on the relationships between theoretical added noises and the thermal occupation of the mechanical mode~\cite{han2021microwave}, we further estimate the thermal occupation of the mechanical mode to be $\nth \sim 22$, and the added noise for microwave-to-optical conversion to be $n_{\text{added}, \text{o}\upmu}  \sim 181$.

\subsection*{Microwave readout and added noise}

\begin{figure*}[tb]
\centering
\includegraphics[scale=0.95]{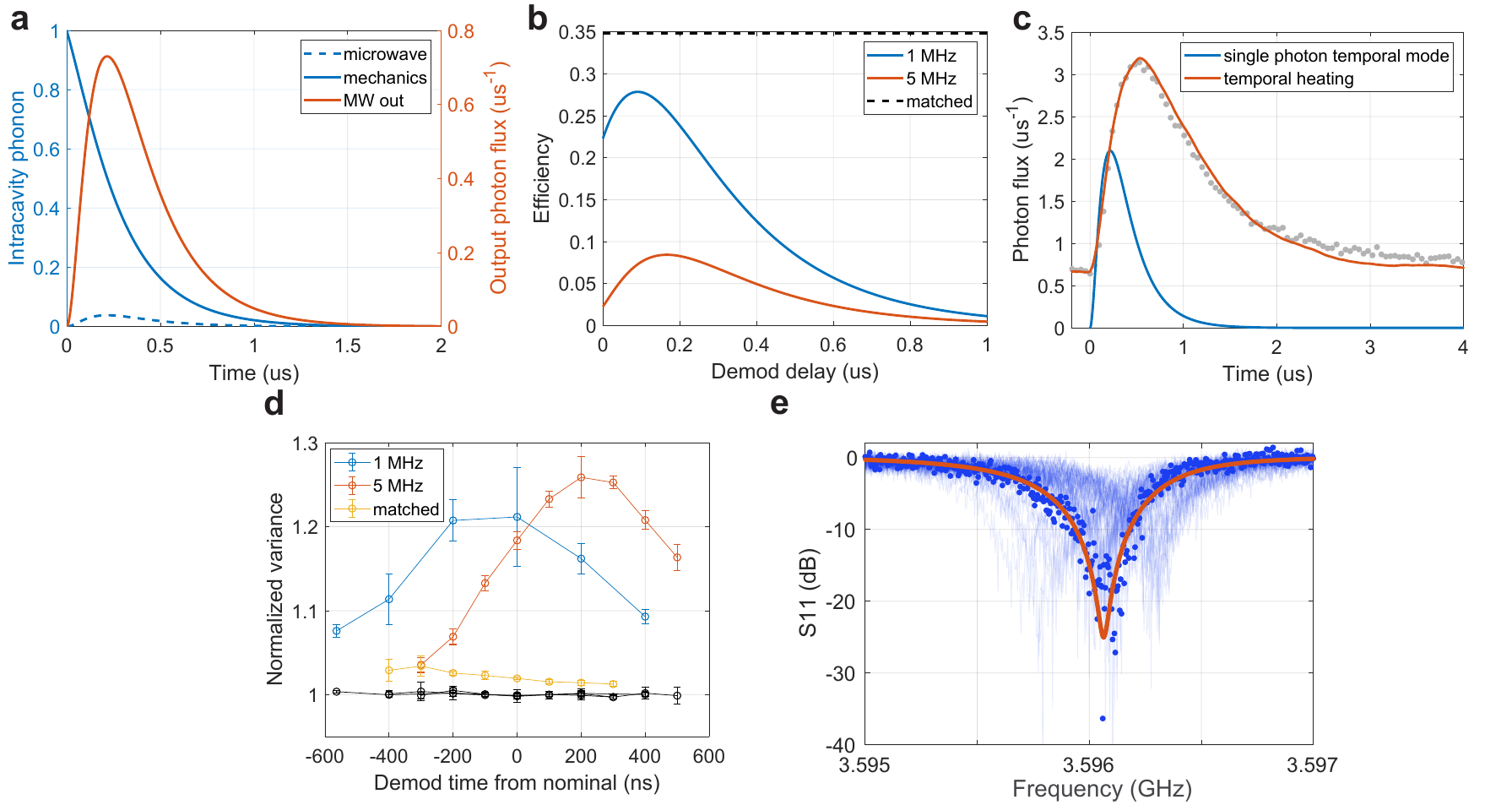}
\caption{\label{SI-Fig:demod} \textbf{Demodulation efficiency and timing.} \textbf{a}, Calculated intracavity phonon and output photon flux versus time of the transducer with one initial phonon using coupled mode theory. \textbf{b}, Calculated mechanics-to-microwave efficiency using different demodulation waveforms. Matched waveform gives the highest possible efficiency and is limited by device internal loss. \textbf{c}, Added noise in the microwave photon. The temporal heating is a theory fit to the time-domain heating measurement, assuming a bath with exponentially decaying thermal noise excited by the laser pulse. The temporal mode of the single photon is shown in comparison. The measured temporal heating is shown as grey dots. \textbf{d}, Measurement of the best demodulation timing for different demodulation waveforms. Efficiency of the demodulation is measured by the relative variance of post-selected state IQ data from single photon detection events versus the thermal state IQ, explained in detail in Methods. Black curves are control calculations using randomly selected IQ data instead of post-selected. \textbf{e}, Microwave $S_{11}$ with no optical pump. $50$ measurements, taken in quick succession of each other are plotted (blue) where one of them is highlighted for clarity (blue dots). The red curve shows the fit result from coupled-mode theory. For some traces the mechanical mode is undercoupled while for others it is overcoupled due to fluctuations of the intrinsic mechanical loss rate. The frequency is also stochastically jumping around. These effects are possibly due to two-level systems (TLS).}
\end{figure*}

The microwave readout uses a phase-insensitive amplifier with $ \Gm \gg 1$, necessarily adding noise which we represent with $\op{h}{}$,
\begin{equation}
    \op S{} = \sqrt{\Gm} \cout +\sqrt{\Gm-1} \opd h{} \approx \sqrt{\Gm} (\cout + \opd h{}),
\end{equation}
where $\op c{\text{out}}$ is the microwave output operator. The output operator is related to the mechanical mode operator by the external coupling efficiency  $\etamu$. Note that $\etamu$ is different from the external coupling efficiency of the microwave resonator $\eta_\upmu$.  Assuming perfect demodulation, the temporal microwave mode detected from the device has a ladder operator given by
\begin{equation}
    \cout = \sqrt{\etamu} \op b{} + \sqrt{1-\etamu} \op b {\text{n}},
\end{equation}
where $ \op b{\text{n}} $ describes noise added from other degrees of freedom because of the device inefficiency, and follows the relations $ \braket{ \op b{\text{n}} \opd b{\text{n}}} = n_\text{n} + 1$ and $ \braket{ \opd b{\text{n}} \op b{\text{n}}} = n_\text{n}$.

Noise is further added by the microwave readout chain. More specifically, we characterize the microwave readout by gain $\Gm$, measurement noise $ n_{\text{m}}$ referred to the TWPA input, and the demodulation efficiency $ \eta_\text{d}$,
\begin{eqnarray}
    \op I{} &=& \op S{} + \opd S{}\nonumber\\
    &=& \sqrt{\etademod \etamu \Gm} \op X{} \nonumber\\
    &&+ \sqrt{\etademod (1-\etamu) \Gm} \op X {\text{n}} + \sqrt{\Gm} \op X{\text{m}},
\end{eqnarray}
where $\op X{(\text n)} = \op b {(\text n)} + \opd b{(\text n)}$. The measurement added noise is assumed to be broadband and is independent of the demodulation. $ X_{\text{n(m)}} $ follows the noise statistics $ \braket{X_\text{n(m)}^2} = 2n_\text{n(m)} + 1 $.

When the mechanical mode is in a thermal state with mean phonon number $\nth$, the variance of the demodulated IQ data is
\begin{eqnarray}
\label{eq:Isq}
    \braket{I^2} &=& \etademod \etamu \Gm (2\nth + 1) + \etademod (1-\etamu)\Gm (2n_\text{n} +1) \nonumber\\
    &&+ \Gm (2n_\text{m} +1)\nonumber\\
    &=& \etademod \etamu \Gm (2 (\nth  + \nex) + 2),
\end{eqnarray}
where $n_\text{n}$ is thermal noise from other degrees of freedom, including heating from the optical pump pulse, emitted into the microwave channel, and $n_\text{m}$ is excess noise from the microwave amplifiers. We have lumped all excess noise into $ \nex $,
\begin{eqnarray}
    \nex &=& \frac{1-\etamu}{\etamu} n_\text{n} + \frac{1}{\etademod \etamu} n_\text{m} \nonumber \\
    && + \frac{1}{2}\left( \frac{1-\etamu}{\etamu} + \frac{1}{\etademod \etamu} - 1  \right).
\end{eqnarray}

For ideal quadrature-phase measurement, $\etamu = \etademod = 1$ and $n_\text{m} = 0$, and a minimal noise of $1/2$ is added, giving us the phase space distribution as the Husimi Q representation of the state. Note that this minimal added noise is not included in our definition of excess noise $\nex$, and $\nex = 0$ for ideal quadrature-phase measurement.

When a phonon is added to the state via post-selection with single photon detection, the mean phonon number is $ n_\text{th, PS} = 2\nth + 1 $. As a result,
\begin{eqnarray}
    \braket{I^2}|_\text{PS} &=& \braket{I^2} + \etademod \etamu \Gm (2\nth + 2),\\
    \frac{\braket{I^2}|_\text{PS}}{\braket{I^2}} & = & 1 + \frac{\nth + 1}{\nth + \nex + 1}.
\end{eqnarray}
In the actual experiment, there is a non-zero dark count rate of $70\pm\SI{10}{\hertz}$ on the SPD, resulting in a heralding efficiency $\etaherald \approx 85\% $, defined as the fraction of the total counts that are actually from the sideband photons. The normalized post-selected variance is then given by
\begin{equation}
\label{eq:Isq-ratio}
    \frac{\braket{I^2}|_\text{PS}}{\braket{I^2}}  =  1 + \etaherald \frac{\nth + 1}{\nth + \nex + 1}.
\end{equation}

To minimize added noise $n_\text{n}$ in the microwave tomography measurement, matched filtering is desired and can be realized during the digital demodulation~\cite{eichler2011experimental, patel2021room}. The optimal filter shape is given by the time domain waveform of the outgoing photon, which has the same time-domain waveform as the classical solution of the propagating microwave field $ A(t) $. We solve the coupled-mode theory (CMT) to calculate the propagating microwave field with one initial phonon in the mechanical mode, as shown in \ref{SI-Fig:demod}a. The intracavity photon number of the mechanical and microwave modes are also shown for comparison.

We have attempted demodulation with different filtering including the matched filtering with the numerical waveform $f_\text{demod}(t) = A(t)$ calculated from CMT, and exponential waveforms $ f_\text{d}(t) =  \sqrt{\kappa_\text{d}} \exp(-\kappa_\text{d} t/2)\theta(t) $ with decay rate of $ \kappa_\text{d}/2\pi = \SI{1}{\mega\hertz} $ and \SI{5}{\mega\hertz}. $\theta(t)$ is the Heaviside function. The matched filter is plotted as a dashed black line to show the theoretical maximal efficiency limited by the device parameters. \ref{SI-Fig:demod}b shows the theoretical measurement efficiency $ \eta_\text{m}(t) \equiv |\int d\tau A^*(\tau) f_\text{d}(\tau- t) |^2 $, where $t$ is the demodulation delay.  The matched waveform gives the highest possible efficiency $ \etamu$ which is the device efficiency, and is limited by device internal loss. \ref{SI-Fig:demod}c shows the measured relative change of the IQ variance before and after post-selection as a figure of merit for the state tomography measurement~\cite{patel2021room}. A higher initial thermal phonon occupation both increases the sideband count rate, and makes the post-selected state more distinguishable with the same excess noise from the measurement, since the mean phonon number roughly doubles after the phonon addition event~\cite{patel2021room}. Therefore, we apply a heating pulse to the transducer \SI{140}{\nano\second} before the heralding pulse to increase the initial thermal phonon occupation to $\sim 10$. We find that the two exponential filters give a more visible change in variance, and thus lower excess noise. We suspect this is due to two-level system induced fluctuation of the mechanical frequency, as shown in \ref{SI-Fig:demod}e with microwave reflection measurement under no optical pump. As a result, the demodulation in the heralding experiment is conducted with a \SI{5}{\mega\hertz} exponential filter, and the theoretical total measurement efficiency is $\approx 8\%$, where the demod efficiency $\etademod \approx 24\% $ and the device efficiency $\etamu \approx 35\%$. This is larger than the magnitude of the frequency fluctuations and therefore masks their effect at the cost of lowered efficiency. Characterization and improvement of the frequency fluctuation will be the subject of future study.

In order to calibrate the gain $\Gm$ and excess noise $n_\text{m}$, two different states of the microwave field are required. We carry out two types of calibration. First, we use the laser heating to generate two thermal states with different thermal occupations. For the thermal state with higher $\nth$, a heating pulse is applied \SI{140}{\nano\second} prior to the probe pulse, where the sideband asymmetry from the probe pulse is used to obtain the thermal occupation of the mechanical mode $n_\text{th,high} = 5.9\pm 2$. For the thermal state with lower $\nth$, we use a repetition of \SI{10}{\micro\second} and no heating pulse to obtain $n_\text{th, low} = 0.56 \pm 0.2$. The demodulation is aligned to the probe pulse, and we measure the IQ variance from these two different thermal states. From Eq.~\ref{eq:Isq}, we calculate $ \nex = 18 \pm 8$. However, we find that the device parameters are varying under different optical powers, which results in different $\etamu$. A \SI{10}{\percent} change in $\etamu$ could lead to a factor of $2$ difference in $\nex$. Alternatively, $\nex$ can be calculated using the thermal and photon-added state as shown in Eq.~\ref{eq:Isq-ratio}. For the data shown in Fig.~\ref{Fig3:herald} of the main text, we find $ \nex = 39 \pm 6$. 

To better understand the potential of the microwave field from the transducer, it is important to estimate the added noise $n_\text{n}$ in it. Noise in the microwave field is not directly measurable with imperfect detection. Nevertheless, we carry out the estimation with two different methods. In addition to the pre- and post-selected datasets, we further take a control measurement every $500$ experimental runs, where we execute the same demodulation with the same timing except that no optical pump pulse is sent to the device. The system can be approximated to be in the initial thermal state with $\nth = 0.68$, independently measured by optical sideband asymmetry, and no heating or microwave photon is generated. As a result, the variance becomes
\begin{equation}
    \braket{I^2}|_\text{th} = \etademod \Gm (2 \nth + 1) + \Gm (2 n_\text{m} + 1).
\end{equation}
Using $\etademod$ and $\etamu$ calculated from theory, the measured $ \braket{I^2}$, $ \braket{I^2}|_\text{PS}$ and $\braket{I^2}|_\text{th}$ allow us to calculate $ n_\text{n} = 1.9~\pm~0.4 $ and $n_\text{m} = 2.4~\pm~0.4$.

Alternatively, we could estimate the added noise in the microwave field from the overlap between the measured temporal heating and the temporal mode of the microwave photon. We fit the measured temporal heating assuming the mechanical mode is coupled to a thermal bath with an exponentially decaying population excited by the pump pulse. Noise in the output microwave signal is calculated with semi-classical Monte Carlo simulation of an ensemble of 3000 instances. \ref{SI-Fig:demod}c shows the temporal heating together with the single photon temporal mode. Overlap between the heating and single photon temporal mode gives $ n_\text{n} = 1.3\pm 0.2$. The uncertainty mostly comes from the sideband asymmetry measurement of the initial $\nth$.

\subsection*{Measurement setup}

\begin{figure*}[tb]
\centering
\includegraphics[scale=0.6]{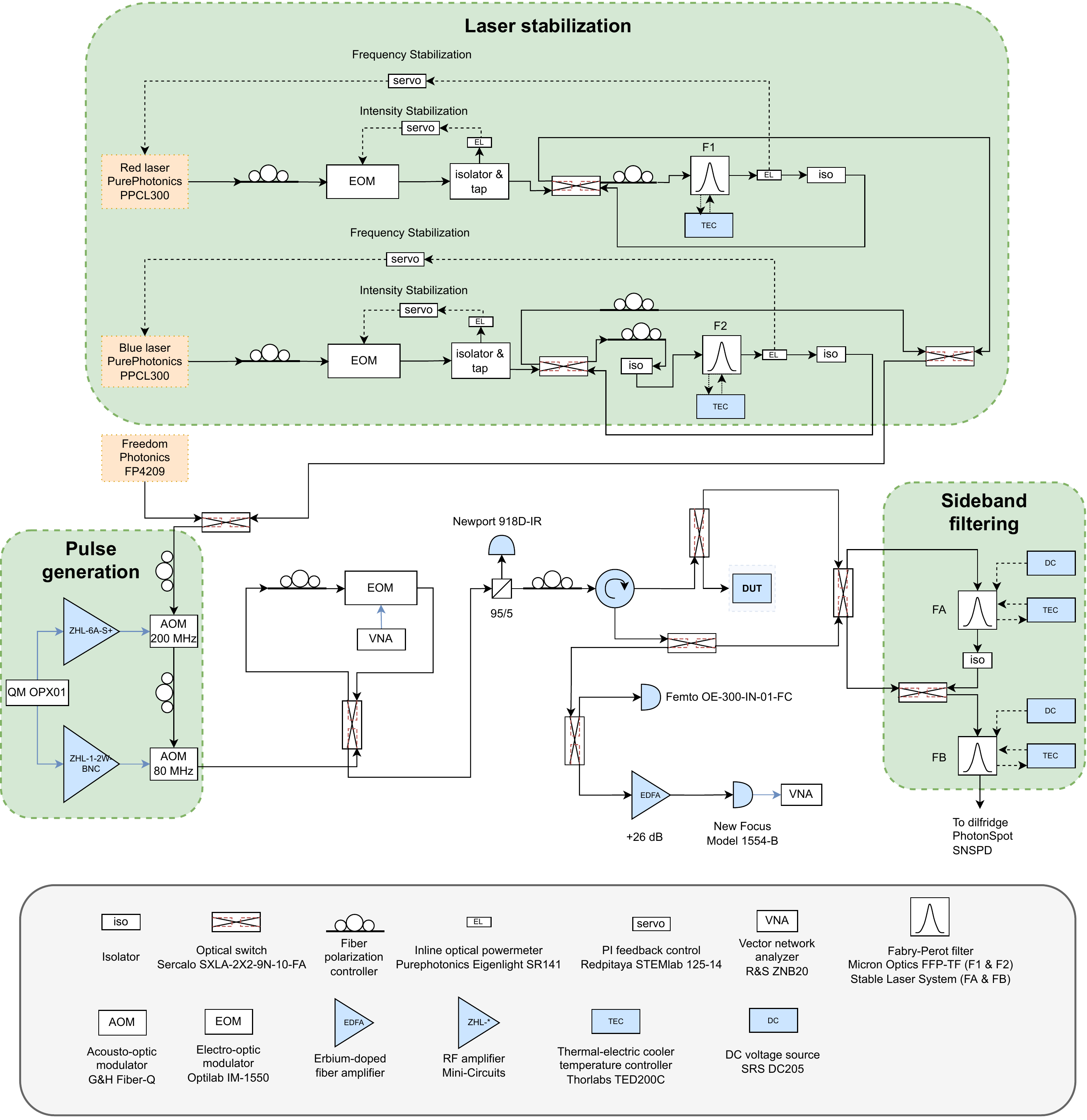}
\caption{\label{SI-Fig:optical-setup} \textbf{Optical setup.}}
\end{figure*}

\ref{SI-Fig:optical-setup} shows the optical setup used in this work. Two tunable external cavity diode lasers (PurePhotonics PPCL300) are first intensity-stabilized with electro-optic modulators, and then frequency-stabilized using temperature-stabilized fiber Fabry-Pérot filters (F1 and F2). The filters also suppress the laser phase noise at the converter frequencies. A fast wavelength-scanning laser (Freedom Photonics FP4209) is used for fiber-to-chip coupling optimization. Two acousto-optic modulators (AOMs) are simultaneously pulsed to generate the optical pump pulse with high on-off ratio ($> \SI{90}{\decibel}$). The duration of the pulse is limited by the rise-fall time of the \SI{200}{\mega\hertz} AOM to be $\gtrsim \SI{20}{\nano\second}$.  Multiple MEMS optical switches are implemented to route the input light to either the transducer or the filter cavities FA and FB for coarse tuning. The filter cavities have \SI{15}{\mega\hertz} bandwidth and \SI{14}{\giga\hertz} free spectral range, and provide \SI{90}{\decibel} pump suppression, while dispersion at the filter resonance delays the sideband photons by an extra $\sim \SI{40}{\nano\second} $ compared to the feed-through transmitted pump photons. This allows us to further separate the pump and sideband photon counts in time domain. The reflected light from the device can be routed to one of the photodetectors for the EIT measurement or microwave-to-optical conversion measurement, or to the filter cavities for single photon detection on the sideband photons.

\begin{figure*}[tb]
\centering
\includegraphics[scale=0.9]{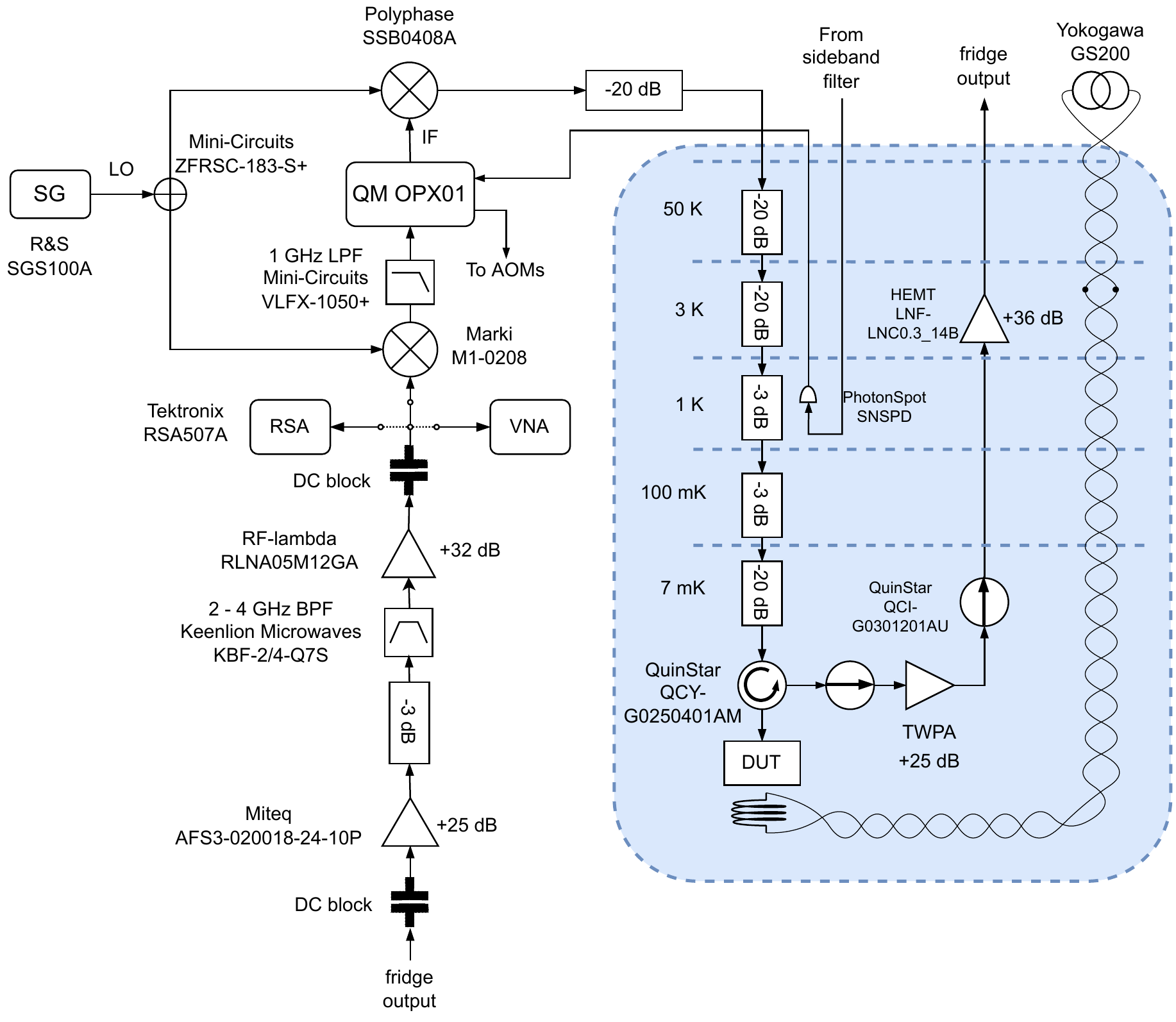}
\caption{\label{SI-Fig:microwave-setup} \textbf{Microwave setup.}}
\end{figure*}

We show the microwave setup in \ref{SI-Fig:microwave-setup}. An intermediate frequency (IF) of \SI{125}{\mega\hertz} is used from the Quantum Machine (QM OPX01) and upconverted for the microwave input. Proper attenuations on the microwave input line guarantee the input microwave thermal noise to be less than $0.01$. The microwave signal from the transducer is first amplified by a traveling-wave parametric amplifier (TWPA) with the dispersive feature around \SI{6.42}{\giga\hertz} and pumped at \SI{5.027}{\giga\hertz} (not shown) to maximize the gain at the transducer frequency. Two broadband isolators ($3-\SI{12}{\giga\hertz}$) are installed before and after the TWPA to minimize reflection within its gain bandwidth. The signal is then further amplified by a high electron mobility transistor (HEMT) and two room temperature low noise amplifiers, and measured by either the real-time spectrum analyzer (RSA), the vector network analyzer (VNA), or downconverted and digitized on the QM. Temporal delay in the optical setup is longer than in the microwave setup, and the microwave signal from the transducer arrives at the QM prior to the voltage pulse from the SPD. As a result, the demodulation is always executed first, and then stored differently in real-time, conditioned on the SPD event.

\section*{Data availability}
The data that support the findings of this study are available from the corresponding author upon reasonable request.

\section*{Acknowledgment}
 W.J. and F.M.M. would like to thank Christopher J. Sarabalis and Haonan Xiong for helpful discussions. A-H.S.N. acknowledges useful discussions with Oskar Painter, Cindy Regal, Konrad Lehnert, Martin Fejer, and Simon Groeblacher. The authors would like to thank Kevin K.S. Multani, Agnetta Y. Cleland, Oliver A. Hitchcock, Carsten Langrock, and Matthew P. Maksymowych for fabrication assistance, Kevin A. Villegas Rosales and Niv Drucker at Quantum Machines and Yudan Guo for technical support, and Kyle Serniak and William D. Oliver at MIT Lincoln Laboratory for providing the TWPA. This work was primarily supported by the U.S. Army Research Office (ARO) Cross-Quantum Systems Science \& Technology (CQTS) program (Grant No. W911NF-18-1-0103), the National Science Foundation CAREER award No.~ECCS-1941826, the Airforce Office of Scientific Research (AFOSR) (MURI No. FA9550-17-1-0002 led by CUNY), and the David and Lucille Packard Fellowship. Device fabrication was performed at the Stanford Nano Shared Facilities (SNSF) and the Stanford Nanofabrication Facility (SNF), supported by the NSF award ECCS-2026822. A.H.S.-N. acknowledges support via a Sloan Fellowship.  The authors also wish to thank NTT Research and Amazon Web Services Inc. for their financial support. Some of this work was funded by the U.S. Department of Energy through Grant No. DE-AC02-76SF00515 and via the Q-NEXT Center.
 
\section*{Author contributions}
W.J. designed the device with assistance from F.M.M and S.M.. W.J. and F.M.M. fabricated the device assisted by S.M.. W.J., F.M.M. and R.V.L. developed the fabrication process. W.J. and F.M.M measured the device with assistance from S.M.. R.N.P., T.P.M., J.D.W. and A.H.S.-N. provided assistance with the measurement setup. W.J., F.M.M. and A.H.S.-N. wrote the manuscript with input from all authors. A.H.S.-N. supervised the project.

\section*{Competing interests}
A.H.S.-N. is an Amazon Scholar. The other authors declare no competing interests.

\onecolumngrid

\section*{Supplementary information}

\renewcommand{\figurename}{}
\renewcommand{\thesection}{Supplementary Note \arabic{section}}
\renewcommand{\thefigure}{Supplementary Figure \arabic{figure}}

\setcounter{figure}{0}

\renewcommand{\tablename}{Extended Data Table}
\setcounter{table}{0}

\section{Device design}

The optomechanical design is similar to previous piezo-optomechanical transducer works~\cite{peairs2020continuous, mirhosseini2020superconducting}. We design two different types of mirror unit cells, one optimized with both optical and mechanical bandgaps around the frequency of the localized modes for the optical and mechanical polarizations of interest, and one with only the optical bandgap while still allowing the mechanical mode to propagate. We first optimize the optomechanical crystal with a fully symmetric design using the first type of unitcell on both ends of the OMC, and optimize the defect cell geometries to maximize the product of the radiation-limited optical quality factor and the optomechanical coupling rate $\go$. A cubic transition between the defect unit cell and the mirror unit cell forms the smooth confinement of the optical and mechanical mode. Unit cells on one side of the OMC are then replaced to be the second type, enabling the hybridization between the mechanical mode in the OMC and the hybrid Si-LN piezoelectric mode.

It is important to maximize the hybridization between the OMC mechanical mode and the Si-LN piezoelectric mode, so that the device is tolerant to fabrication variations. We note that the leaky mechanical mode from the OMC is mostly longitudinal, while the strong piezoelectric mode in the Si-LN structure is mostly in horizontal shear (SH) polarization. We use a $45^\circ$ rotation between the OMC and the Si-LN structure to align the polarization between the longitudinal mode and SH mode, and achieve a hybridization with coupling strength $J/2\pi \approx \SI{42}{\mega\hertz}$. Similar to Ref.~\cite{arrangoiz2019resolving}, we adopt $X$-cut thin-film LN to utilize the strong piezoelectricity of the SH mode, where its largest piezoelectric component $d_{24} \approx \SI{70}{\pico\meter\per\volt}$ couples $YZ$ shear to electric field along the crystal $Y$ axis. The aluminum electrodes are patterned to be parallel to the crystal $Z$ axis so that the electric field between them are parallel to $Y$. To prevent the mechanical mode from leaking into the substrate, we implement extra one dimensional phononic shields (1DPS) with a full bandgap around the mechanical frequency. The 1DPS is optimized separately for anchoring the Si-LN piezoelectric structure to take into account the aluminum electrodes.

\section{Device fabrication}

\begin{figure*}[tb]
\centering
\includegraphics[scale=0.9]{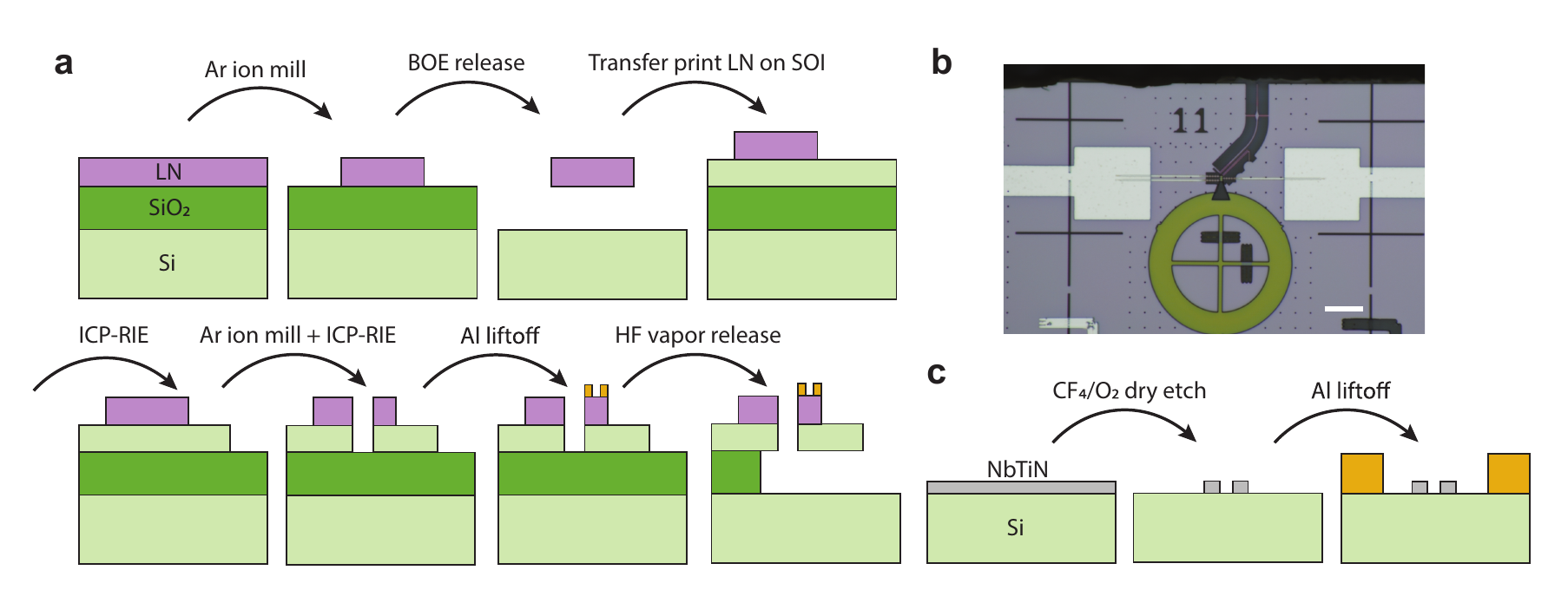}
\caption{\label{SI-Fig:fab} \textbf{Device fabrication.} \textbf{a}, Fabrication process of the transducer chip. \textbf{b}, Optical image of the transducer device after the full fabrication process. The edge of the chip is diced to expose the on-chip optical coupler. \textbf{c}, Fabrication process of the microwave chip.}
\end{figure*}

Patterning of thin film LN has been done with a wide variety of underlying substrates such as silicon, silica and sapphire. However, physical etching of LN is not directly compatible with the LN-on-SOI platform since the etch damages the surface of the silicon thin film and leads to significantly reduced optical quality factors. To avoid this issue, we pattern a thin film of LN on a separate LN-on-insulator sample and transfer print~\cite{meitl2006transfer} it onto a pristine SOI chip.
The fabrication process is summarized in \ref{SI-Fig:fab}. We start with 400 nm thick MgO-doped X-cut LN on a \SI{3}{\micro\meter} thick thermal silica buffer layer sitting on a \SI{400}{\micro\meter} thick silicon handle (commercially available from NanoLN). The LN film is first thinned to \SI{290}{\nano\meter} with argon ion milling. The LN transducer and alignment marks are defined with electron beam lithography (EBL) using hydrogen silsesquioxane (HSQ) as the resist. The pattern is transferred to LN with a \SI{265}{\nano\meter} deep etch using argon ion milling. The HSQ is stripped in buffered oxide etchant (BOE, 6:1) and the LN is then cleaned in dilute hydrofluoric acid (HF, \SI{5}{\percent}). To etch the remaining LN slab and expose the silica, we carry out another \SI{40}{\nano\meter} deep LN etch with blanket argon ion milling, followed by a clean in a piranha solution (3:1 $98\% $ sulfuric acid and $40\% $ hydrogen peroxide). The LN is released by etching the silica in 6:1 BOE for $22$ minutes which leads to a \SI{3}{\micro\meter} undercut of the thermal oxide. A piece of polydimethylsiloxane (PDMS) attached to a glass-slide which is mounted on a set of translational stages is lowered until it fully covers the LN sample. Due to its viscoelasticity~\cite{meitl2006transfer}, when rapidly lifting up the PDMS away from the sample, released LN patterns break off from their anchors and stick to the PDMS. We then make the PDMS with LN touch down onto a SOI piece (\SI{220}{\nano\meter} thick silicon device layer, \SI{3}{\micro\meter} thick buried oxide, \SI{725}{\micro\meter} silicon handle) that has been cleaned in a piranha solution. By slowly peeling off the PDMS from the SOI substrate, the LN patterns are released from the PDMS and adhere to the silicon. We could simultaneously transfer print more than 100 devices, and typically achieve a $>  \SI{95}{\percent}$ transfer print yield. Subsequently, the chip is annealed for $8$ hours at \SI{500}{\celsius} to improve the LN-Si adhesion. This is followed by a clean in piranha solution. The transfer print causes a $\sim 0.1\%$ relative displacement between separated patterns. We attach alignment marks within \SI{30}{\micro\meter} of every device to enable sub-\SI{30}{\nano\meter} alignment error on EBL masks after the transfer print.

We pattern the silicon with aligned EBL and do an inductively coupled plasma (ICP) - reactive ion etch (RIE) using a $\text{Cl}_2$ and HBr gas chemistry. The device is still attached to the transfer printed alignment mark which would cause mechanical loss. We therefore do another aligned EBL to pattern a small window where we argon ion mill the LN followed by ICP-RIE to etch through the silicon device layer. Subsequently, we expose the oxide in the areas of the chip that do not have any devices by etching the silicon device layer with ICP-RIE after having defined etch windows with photolithography. 
The sample is cleaned a final time in a piranha solution before metalization. The transducer electrodes are patterned with aligned EBL. We evaporate \SI{55}{\nano\meter} aluminum at three different angles ($-66^\circ, 0^\circ, 66^\circ$) to help the electrodes climb the LN from the silicon, and lift-off in N-Methyl-2-pyrrolidone (NMP) at \SI{80}{\celsius}. The wirebond pads are defined with photolithography, and we evaporate $\SI{250}{\nano\meter}$ aluminum followed by lift-off in NMP at \SI{80}{\celsius}. Finally, the chip is diced close to the edge coupler and the devices are released in HF vapor. To accommodate fabrication uncertainties, the optomechanical crystals and the separation of the piezoelectric resonator electrodes in different transducer devices are scaled by $\pm 2\%$ and $ \pm 5\% $ respectively.

The microwave resonator is fabricated on a separate die from the transducer to simplify fabrication and potentially reduce optically induced quasiparticle generation~\cite{mirhosseini2020superconducting} by vertically offsetting the resonator from the optical fiber. We start with a \SI{10}{\nano\meter} thin layer of niobium titanium nitride (NbTiN) on high-resistivity silicon. We use EBL to define the meander in HSQ resist and etch the NbTiN using RIE with a $\text{CF}_4$ and $\text{O}_2$ gas chemistry. The HSQ is stripped in 6:1 BOE. Finally, a \SI{250}{\nano\meter} thick aluminum ground plane is defined with photolithography followed by evaporation and liftoff.

\section{Room temperature mechanics-to-microwave external coupling measurement}

\begin{figure*}[tb]
\centering
\includegraphics[scale=1.]{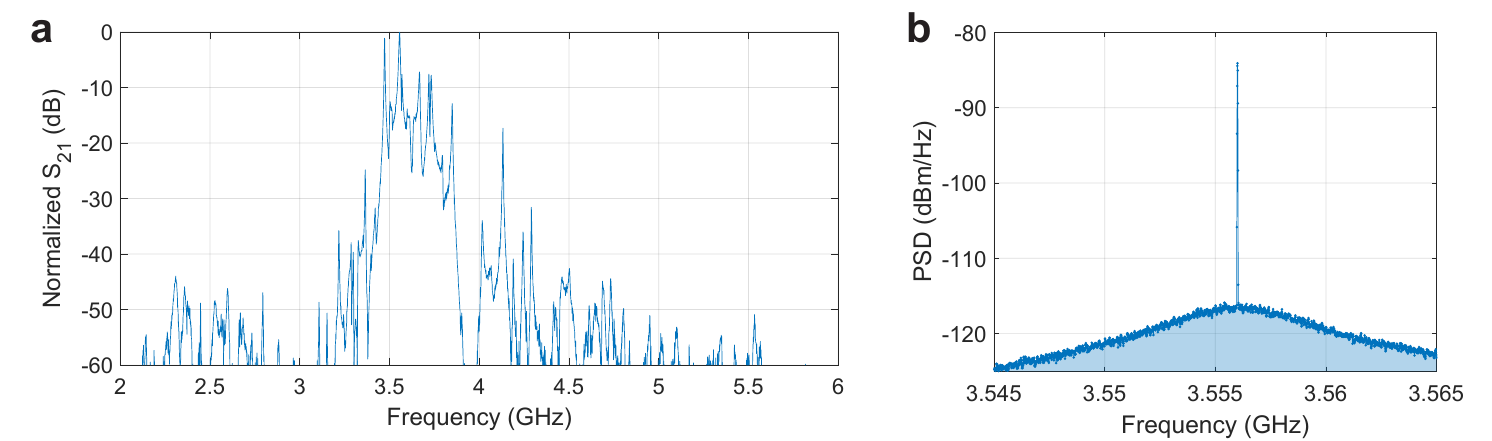}
\caption{\label{SI-Fig:RT-conversion} \textbf{Room-temperature conversion measurement and calibration.} \textbf{a}, Normalized microwave-to-optical conversion S parameter. Multiple mechanical modes are visible due to the high signal-to-noise ratio of the measurement. Two mechanical modes are strongly hybridized with the OMC mechanical mode and show much higher conversion efficiency than other modes. \textbf{b}, Calibration of the mechanics-to-microwave external coupling. The thermal-mechanical noise power (shaded area) is used to calibrate the optomechanical readout of the intracavity phonon number.}
\end{figure*}

The fabricated transducers can be measured at room temperature before integrating with the microwave chip, especially for characterizing the mechanics-to-microwave external coupling and optical quality factor. This allows us to select the best performing transducer for packaging and cooldown. Similar to the method described in Ref.~\cite{jiang2019lithium}, a microwave probe (GGB Industries, Picoprobe model 40A-GSG) is brought in contact with the on-chip electrodes of the piezo-optomechanical transducer. A red-detuned laser is used to enable the microwave-to-optical conversion and readout the converted sideband photon (\ref{SI-Fig:RT-conversion}). The coherent microwave input signal is then fixed at the peak conversion frequency, on resonance with one of the mechanical modes, and gives rise to a coherent intracavity phonon population
\begin{equation}
    n_\text{coh} = \frac{4\gammamu\dot{N}}{\gamma^2},
\end{equation}
where $\gammamu$ is the external coupling rate between the mechanical mode and the \SI{50}{\ohm} microwave line, $\gamma$ is the total mechanical linewidth, and $\dot N$ is the input microwave photon flux. As shown in ~\ref{SI-Fig:RT-conversion}, both the coherent and the thermal phonons are converted to sideband photons, and measured on the microwave power spectrum from a fast photodetector output. To calculate the thermal phonon number and calibrate the optical readout, the device temperature is assumed to be \SI{295}{\kelvin} . The coherent phonon number $n_\text{coh}$ and the external coupling rate $\gammamu$ can then be calculated.

\section{Piezoelectric coupling via cross-chip wirebond}

\begin{figure*}[tb]
\centering
\includegraphics[scale=0.58]{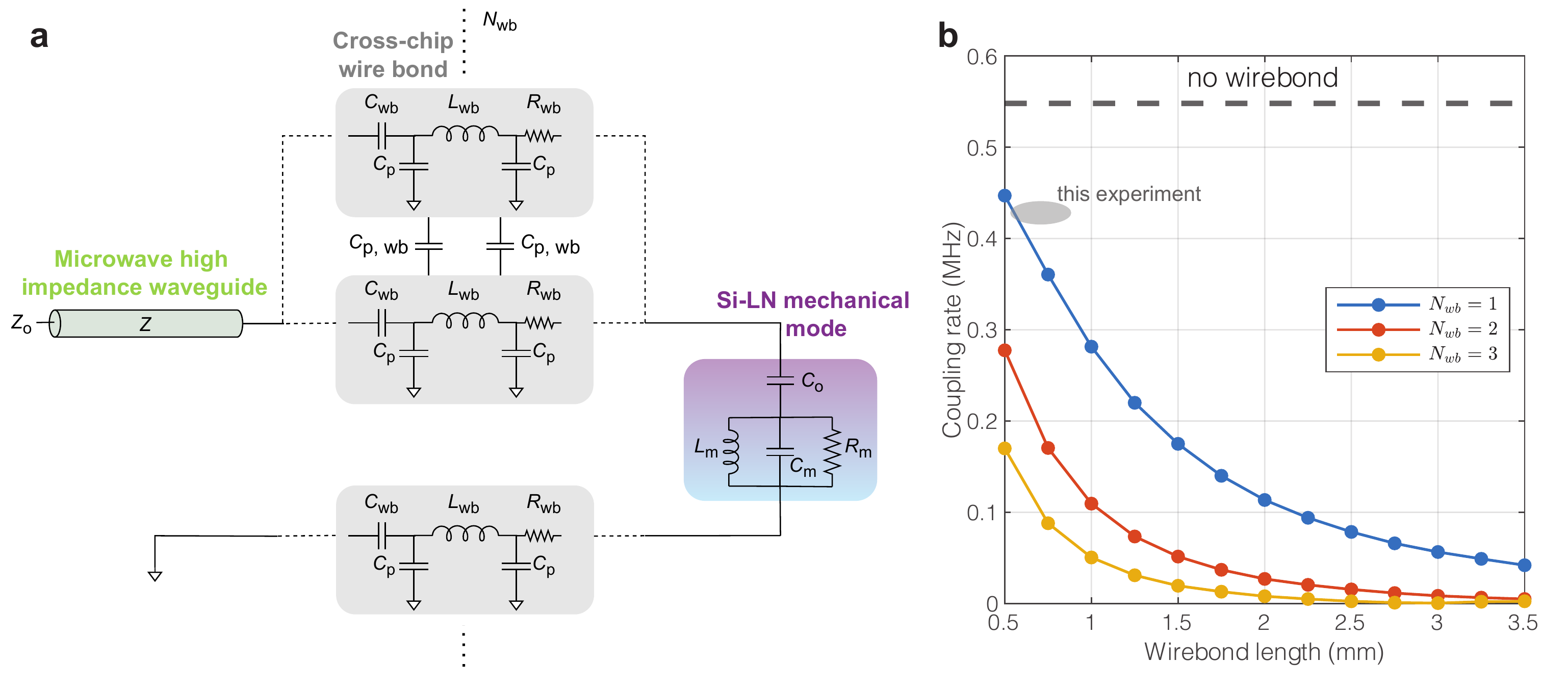}
\caption{\label{SI-Fig:xcwb} \textbf{Cross-chip wire bond model.} \textbf{a}, Lumped element circuit model of the cross-chip wirebonds between the Si-LN mechanical mode of the transducer chip and the high-impedance waveguide mode of the microwave chip. The Si-LN mechanical mode is modeled as a $RLC$ circuit (with parameters $R_{\text m}$, $L_{\text{m}}$, $C_{\text m}$, and $C_{o}$) and the microwave waveguide is modeled as a transmission line (with a characteristic impedance of $Z = \SI{1000}{\ohm}$). The signal and ground cross-chip wirebonds are modeled as series inductance $ L_{\text{wb}}$ with parasitic capacitance to the ground plane $C_{\text p}$ in parallel, a contact capacitance $C_{\text{wb}}$ and a contact resistance $R_{\text{wb}}$ in series. There are $N_{\text{wb}}$ number of wirebonds with mutual capacitance $C_{\text{p,wb}}$ between them. \textbf{b}, Coupling rate between the Si-LN mechanical mode and the microwave waveguide as a function of wirebond length for various values of $N_{\text{wb}}$. A shaded grey region shows the estimated coupling rate and the estimated wirebond length in this experiment.}
\end{figure*}

The transducer and the microwave resonator are fabricated on separate chips and integrated together via cross-chip wirebonding. To understand the impact of wirebonds on the coupling rate between the two, we consider a lumped element circuit model of the wirebonds between the Si-LN mechanical mode of the transducer chip and the high-impedance waveguide mode of the microwave chip, as shown in \ref{SI-Fig:xcwb}. The Si-LN mechanical mode is modeled as a $RLC$ circuit with parameters $L_{\text{m}} = \SI{195.6}{\nano\henry}$, $C_{\text m} = \SI{10.01}{\femto\farad} $, $R_{\text m} = \SI{45.4}{\mega\ohm}$, and $C_{o} = \SI{0.11}{\femto\farad}$, estimated from a combination of finite-element simulation and room temperature piezo-optomechanical measurements. The microwave waveguide is modeled as a transmission line (FSR =  \SI{110}{\mega\hertz}, $Z = \SI{1000}{\ohm}$). We model the signal and ground cross-chip wirebonds as series inductance $L_{\text{wb}}$ with parasitic capacitance to the ground plane $C_{\text p}$ in parallel, and a contact capacitance $C_{\text{wb}}$ and a contact resistance $R_{\text{wb}}$ in series. The wirebonds have mutual capacitance of $C_{\text{p,wb}}$ between them. For a $\SI{25}{\micro\meter}$ diameter wire, the inductance roughly scales as $L_{\text{wb}}$ $\sim\SI{1}{\nano\henry}/\SI{}{\milli\meter}$ \cite{Wenner2011, huang2021microwave}. The remaining wirebond parameters are estimated using finite-element electrostatic analysis with nominal values of $C_{\text{p}}$ $\sim\SI{20}{\femto\farad}$, $C_{\text{wb}}$ $\sim\SI{20}{\pico\farad}$, $R_{\text{wb}}$ $\sim\SI{0.4}{\ohm}$~\cite{zhong2021deterministic}, and  $C_{\text{p,wb}}$ $\sim\SI{12}{\femto\farad}/\SI{}{\milli\meter}$. The model exhibits strong reduction in the coupling rate with increasing wirebond length and increasing number of wirebonds as seen in the plot in \ref{SI-Fig:xcwb}. This is due to the increase in parasitic capacitance of the wires to the ground plane and increase in mutual capacitance between the signal and ground wirebonds. Hence, to maximize the coupling rate, a single signal wirebond with length as short as possible is desirable. 

We first considered placing the two chips next to each other with the optical edge coupler side of the transducer chip facing away from the microwave chip. However, the shortest possible wirebond in such a case is limited by the net distance between the terminals of the devices on the two chips which turns out to be significant. In this experiment, we utilize a different approach in which the two chips are wirebonded together with the optical edge coupler side of the transducer chip facing the microwave chip. To allow fiber access for optical coupling,  the transducer is then flipped over and stacked vertically, resulting in a shorter wirebond with length on the order of $\SI{0.75}{\milli\meter}$. Two small spacer chips are glued near the edge of the microwave chip with GE varnish to support the transducer chip. Not only does this approach give us a shorter wirebond, it also results in the microwave chip being in a different plane than the optical fiber which likely helps minimize optically induced quasiparticle generation.

\end{document}